\pdfoutput=1

\documentclass[12pt,a4paper]{article}

\usepackage{ifthen} 
\newboolean{pdflatex}
\setboolean{pdflatex}{true} 

\newboolean{articletitles}
\setboolean{articletitles}{true} 

\newboolean{uprightparticles}
\setboolean{uprightparticles}{false} 

\newboolean{inbibliography}
\setboolean{inbibliography}{false} 

\usepackage{array}
\usepackage{amsmath}

\usepackage{microtype}
\usepackage{lineno}  
\usepackage{xspace} 
\usepackage{caption} 

\usepackage{graphicx}  
\usepackage{color}
\usepackage{colortbl}
\graphicspath{{./figs/}} 

\usepackage{amsmath} 
\usepackage{amssymb}
\usepackage{amsfonts}
\usepackage{upgreek} 

\newcommand*\patchAmsMathEnvironmentForLineno[1]{%
\expandafter\let\csname old#1\expandafter\endcsname\csname #1\endcsname
\expandafter\let\csname oldend#1\expandafter\endcsname\csname
end#1\endcsname
 \renewenvironment{#1}%
   {\linenomath\csname old#1\endcsname}%
   {\csname oldend#1\endcsname\endlinenomath}%
}
\newcommand*\patchBothAmsMathEnvironmentsForLineno[1]{%
  \patchAmsMathEnvironmentForLineno{#1}%
  \patchAmsMathEnvironmentForLineno{#1*}%
}
\AtBeginDocument{%
\patchBothAmsMathEnvironmentsForLineno{equation}%
\patchBothAmsMathEnvironmentsForLineno{align}%
\patchBothAmsMathEnvironmentsForLineno{flalign}%
\patchBothAmsMathEnvironmentsForLineno{alignat}%
\patchBothAmsMathEnvironmentsForLineno{gather}%
\patchBothAmsMathEnvironmentsForLineno{multline}%
}

\usepackage{hyperref}    
\usepackage[all]{hypcap} 

\def\lhcb {\mbox{LHCb}\xspace}

\def\babar  {\mbox{BaBar}\xspace}
\def\belle  {\mbox{Belle}\xspace}

\ifthenelse{\boolean{uprightparticles}}%
{
 
 \def\Pgamma      {\ensuremath{\upgamma}\xspace}

 \def\Ppi         {\ensuremath{\uppi}\xspace}

 \def\PDelta      {\ensuremath{\Delta}\xspace}                 
 \def\PXi      {\ensuremath{\Xi}\xspace}                 
 \def\PLambda      {\ensuremath{\Lambda}\xspace}                 
 \def\PSigma      {\ensuremath{\Sigma}\xspace}                 
 \def\POmega      {\ensuremath{\Omega}\xspace}                 
 \def\PUpsilon      {\ensuremath{\Upsilon}\xspace}                 
 
 \def\PB      {\ensuremath{\mathrm{B}}\xspace}                 
                  
 \def\PD      {\ensuremath{\mathrm{D}}\xspace}

 \def\PK      {\ensuremath{\mathrm{K}}\xspace}

 \def\Pb      {\ensuremath{\mathrm{b}}\xspace}                 
 \def\Pc      {\ensuremath{\mathrm{c}}\xspace}

 \def\Pi      {\ensuremath{\mathrm{i}}\xspace}

}
{
 
 \def\Pgamma      {\ensuremath{\gamma}\xspace}

 \def\Ppi         {\ensuremath{\pi}\xspace}

 \mathchardef\PDelta="7101
 \mathchardef\PXi="7104
 \mathchardef\PLambda="7103
 \mathchardef\PSigma="7106
 \mathchardef\POmega="710A
 \mathchardef\PUpsilon="7107
                  
 \def\PB      {\ensuremath{B}\xspace}                 
                  
 \def\PD      {\ensuremath{D}\xspace}

 \def\PK      {\ensuremath{K}\xspace}

 \def\Pb      {\ensuremath{b}\xspace}                 
 \def\Pc      {\ensuremath{c}\xspace}

 \def\Pi      {\ensuremath{i}\xspace}

}

\def\g      {\ensuremath{\Pgamma}\xspace}

\def\cquark    {\ensuremath{\Pc}\xspace}

\def\bquark    {\ensuremath{\Pb}\xspace}

\def\pion  {\ensuremath{\Ppi}\xspace}

\def\pip   {\ensuremath{\pion^+}\xspace}
\def\pim   {\ensuremath{\pion^-}\xspace}
\def\pipm  {\ensuremath{\pion^\pm}\xspace}
\def\pimp  {\ensuremath{\pion^\mp}\xspace}

\def\kaon  {\ensuremath{\PK}\xspace}
  \def\Kbar  {\kern 0.2em\overline{\kern -0.2em \PK}{}\xspace}

\def\Kp    {\ensuremath{\kaon^+}\xspace}
\def\Km    {\ensuremath{\kaon^-}\xspace}
\def\Kpm   {\ensuremath{\kaon^\pm}\xspace}
\def\Kmp   {\ensuremath{\kaon^\mp}\xspace}
\def\KS    {\ensuremath{\kaon^0_{\rm\scriptscriptstyle S}}\xspace}

  \def\Dbar    {\kern 0.2em\overline{\kern -0.2em \PD}{}\xspace}
\def\D       {\ensuremath{\PD}\xspace}

\def\Dz      {\ensuremath{\D^0}\xspace}
\def\Dzb     {\ensuremath{\Dbar^0}\xspace}

\def\B       {\ensuremath{\PB}\xspace}
\def\Bbar    {\ensuremath{\kern 0.18em\overline{\kern -0.18em \PB}{}}\xspace}

\def\Bz      {\ensuremath{\B^0}\xspace}

\def\Bu      {\ensuremath{\B^+}\xspace}
\def\Bub     {\ensuremath{\B^-}\xspace}
\def\Bp      {\ensuremath{\Bu}\xspace}
\def\Bm      {\ensuremath{\Bub}\xspace}
\def\Bpm     {\ensuremath{\B^\pm}\xspace}

  \def\Y#1S{\ensuremath{\PUpsilon{(#1S)}}\xspace}

\def\Lbar {\ensuremath{\kern 0.1em\overline{\kern -0.1em\PLambda}}\xspace}

\def\to                 {\ensuremath{\rightarrow}\xspace}

\def\CP                {\ensuremath{C\!P}\xspace}

\def\AT#1     {\ensuremath{A_{\mathrm{T}}^{#1}}\xspace}           

\def\C#1      {\ensuremath{\mathcal{C}_{#1}}\xspace}                       
\def\Cp#1     {\ensuremath{\mathcal{C}_{#1}^{'}}\xspace}                    
\def\Ceff#1   {\ensuremath{\mathcal{C}_{#1}^{\mathrm{(eff)}}}\xspace}        
\def\Cpeff#1  {\ensuremath{\mathcal{C}_{#1}^{'\mathrm{(eff)}}}\xspace}       
\def\Ope#1    {\ensuremath{\mathcal{O}_{#1}}\xspace}                       
\def\Opep#1   {\ensuremath{\mathcal{O}_{#1}^{'}}\xspace}                    

\newcommand{\tev}{\ifthenelse{\boolean{inbibliography}}{\ensuremath{~T\kern -0.05em eV}\xspace}{\ensuremath{\mathrm{\,Te\kern -0.1em V}}\xspace}}
\newcommand{\gev}{\ensuremath{\mathrm{\,Ge\kern -0.1em V}}\xspace}
\newcommand{\mev}{\ensuremath{\mathrm{\,Me\kern -0.1em V}}\xspace}
\newcommand{\kev}{\ensuremath{\mathrm{\,ke\kern -0.1em V}}\xspace}
\newcommand{\ev}{\ensuremath{\mathrm{\,e\kern -0.1em V}}\xspace}
\newcommand{\gevc}{\ensuremath{{\mathrm{\,Ge\kern -0.1em V\!/}c}}\xspace}
\newcommand{\mevc}{\ensuremath{{\mathrm{\,Me\kern -0.1em V\!/}c}}\xspace}
\newcommand{\gevcc}{\ensuremath{{\mathrm{\,Ge\kern -0.1em V\!/}c^2}}\xspace}
\newcommand{\gevgevcccc}{\ensuremath{{\mathrm{\,Ge\kern -0.1em V^2\!/}c^4}}\xspace}
\newcommand{\mevcc}{\ensuremath{{\mathrm{\,Me\kern -0.1em V\!/}c^2}}\xspace}

\def\mum  {\ensuremath{{\,\upmu\rm m}}\xspace}

\def\invfb   {\ensuremath{\mbox{\,fb}^{-1}}\xspace}

\def\ps   {\ensuremath{{\rm \,ps}}\xspace}

\def\gsim{{~\raise.15em\hbox{$>$}\kern-.85em
          \lower.35em\hbox{$\sim$}~}\xspace}
\def\lsim{{~\raise.15em\hbox{$<$}\kern-.85em
          \lower.35em\hbox{$\sim$}~}\xspace}

\def\pt         {\mbox{$p_{\rm T}$}\xspace}

\def\rad{\ensuremath{\rm \,rad}\xspace}

\def\evtgen     {\mbox{\textsc{EvtGen}}\xspace}

\def\geant      {\mbox{\textsc{Geant4}}\xspace}

\def\photos     {\mbox{\textsc{Photos}}\xspace}

\def\pythia     {\mbox{\textsc{Pythia}}\xspace}

\def\tell1  {TELL1\xspace}
\def\ukl1   {UKL1\xspace}

\usepackage{cite} 
\usepackage{mciteplus}

\begin{document}

\renewcommand{\thefootnote}{\fnsymbol{footnote}}
\setcounter{footnote}{1}


\begin{titlepage}
\pagenumbering{roman}

\vspace*{-1.5cm}
\centerline{\large EUROPEAN ORGANIZATION FOR NUCLEAR RESEARCH (CERN)}
\vspace*{1.5cm}
\hspace*{-0.5cm}
\begin{tabular*}{\linewidth}{lc@{\extracolsep{\fill}}r}
\ifthenelse{\boolean{pdflatex}}
{\vspace*{-2.7cm}\mbox{\!\!\!\includegraphics[width=.14\textwidth]{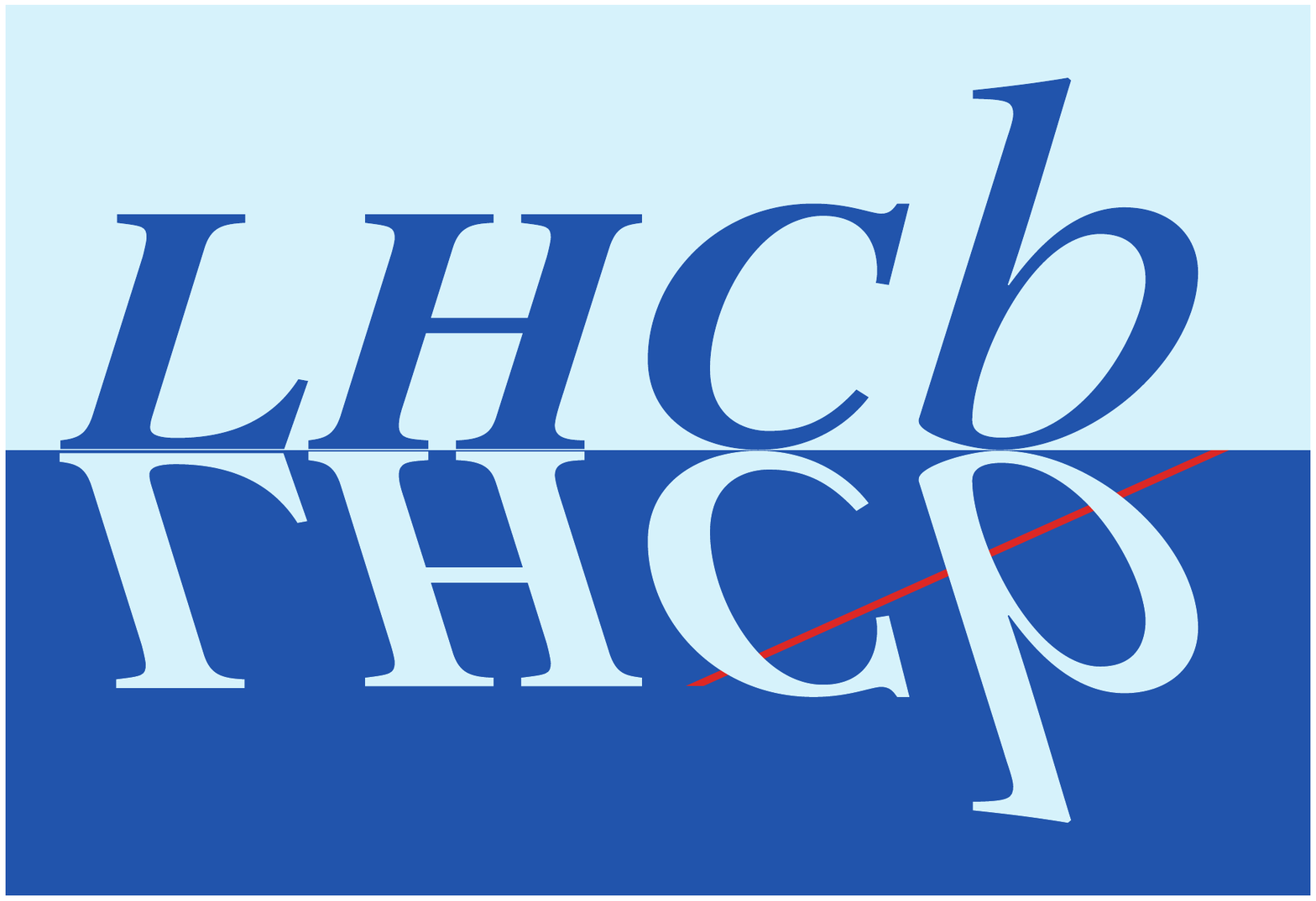}} & &}%
{\vspace*{-1.2cm}\mbox{\!\!\!\includegraphics[width=.12\textwidth]{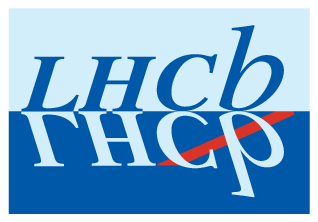}} & &}%
\\
 & & CERN-PH-EP-2014-017 \\  
 & & LHCb-PAPER-2013-068 \\  
 & & 23 May 2014 \\ 
 & & \\
\end{tabular*}

\vspace*{4.0cm}

{\bf\boldmath\huge
\begin{center}
A study of \CP violation in $B^\pm \to D K^\pm$ and $B^\pm \to D\pi^\pm$ decays with $D \to K^0_{\rm S} K^\pm \pi^\mp$ final states

\end{center}
}

\vspace*{2.0cm}

\begin{center}
The LHCb collaboration\footnote{Authors are listed on the following pages.}
\end{center}

\vspace{\fill}

\begin{abstract}
  \noindent
A first study of \CP violation in the decay modes $\Bpm\to[\KS\Kpm\pimp]_Dh^\pm$ and $\Bpm\to[\KS\Kmp\pipm]_Dh^\pm$, where $h$ labels a $K$ or $\pi$ meson and $D$ labels a \Dz or \Dzb meson, is performed. The analysis uses the LHCb data set collected in $pp$ collisions, corresponding to an integrated luminosity of 3\invfb. 
The analysis is sensitive to the \CP-violating CKM phase $\gamma$ through seven observables: one charge asymmetry in each of the four modes and three ratios of the charge-integrated yields. The results are consistent with measurements of $\gamma$ using other decay modes.
 
\end{abstract}
\vspace{-.2cm}
\vspace*{.5cm}

\begin{center}
Published in Phys.~Lett.~B 
\end{center}

\vspace{\fill}

{\footnotesize 
\centerline{\copyright~CERN on behalf of the \lhcb collaboration, license \href{http://creativecommons.org/licenses/by/3.0/}{CC-BY-3.0}.}}
\vspace*{2mm}

\end{titlepage}


\newpage
\setcounter{page}{2}
\mbox{~}
\newpage

\centerline{\large\bf LHCb collaboration}
\begin{flushleft}
\small
R.~Aaij$^{41}$, 
B.~Adeva$^{37}$, 
M.~Adinolfi$^{46}$, 
A.~Affolder$^{52}$, 
Z.~Ajaltouni$^{5}$, 
J.~Albrecht$^{9}$, 
F.~Alessio$^{38}$, 
M.~Alexander$^{51}$, 
S.~Ali$^{41}$, 
G.~Alkhazov$^{30}$
% ...

{\footnotesize \it
% ...
$ ^{5}$Clermont Universit\'{e}, Universit\'{e} Blaise Pascal, CNRS/IN2P3, LPC, Clermont-Ferrand, France\\
% ...
$ ^{9}$Fakult\"{a}t Physik, Technische Universit\"{a}t Dortmund, Dortmund, Germany\\
% ...
$ ^{30}$Petersburg Nuclear Physics Institute (PNPI), Gatchina, Russia\\
% ...
$ ^{37}$Universidad de Santiago de Compostela, Santiago de Compostela, Spain\\
$ ^{38}$European Organization for Nuclear Research (CERN), Geneva, Switzerland\\
% ...
$ ^{41}$Nikhef National Institute for Subatomic Physics, Amsterdam, The Netherlands\\
% ...
$ ^{46}$H.H. Wills Physics Laboratory, University of Bristol, Bristol, United Kingdom\\
% ...
$ ^{51}$School of Physics and Astronomy, University of Glasgow, Glasgow, United Kingdom\\
$ ^{52}$Oliver Lodge Laboratory, University of Liverpool, Liverpool, United Kingdom\\
% ...
}
\end{flushleft}

\cleardoublepage


\renewcommand{\thefootnote}{\arabic{footnote}}
\setcounter{footnote}{0}



\pagestyle{plain} 
\setcounter{page}{1}
\pagenumbering{arabic}


%

\section{Introduction}
\label{sec:Introduction}
A precise measurement of the unitarity triangle angle $\gamma =\arg{\left(-\frac{V_{ud}V_{ub}^*}{V_{cd}V_{cb}^*}\right)}$ is one of the most important tests of the Cabibbo Kobayashi Maskawa (CKM) mechanism. This parameter can be accessed through measurements of observables in decays of charged $B$ mesons to a neutral $D$ meson and a kaon or pion, where $D$ labels a \Dz or \Dzb meson decaying to a particular final state accessible to \Dz and \Dzb. Such decays are sensitive to $\gamma$ through the interference between $b\to c\bar{u}s$ and $b\to u\bar{c}s$ amplitudes. They offer an attractive means to measure $\gamma$ because the effect of physics beyond the Standard Model is expected to be negligible, thus allowing interesting comparisons with other measurements  where such effects could be larger. 

The determination of $\gamma$ using $B^\pm\to D\Kpm$ decays was first proposed for $D$ decays to the \CP-eigenstates $\Kp\Km$ and $\pip\pim$ (so-called ``GLW'' analysis)~\cite{Gronau:1990ra,Gronau:1991dp}. Subsequently, the analysis of the $\Kpm\pimp$ final state was proposed (named ``ADS'')~\cite{Atwood:1996ci,Atwood:2000ck}, where the suppression between the colour favoured $\Bm\to\Dz\Km$ and suppressed $\Bm\to\Dzb\Km$ decays is compensated by the CKM suppression of the $\Dz\to\Kp\pim$ decay relative to $\Dzb\to\Kp\pim$, resulting in large  interference.  The LHCb collaboration has published the two-body ADS and GLW analyses~\cite{LHCb-PAPER-2012-001}, the Dalitz analysis of the decay $B^\pm\to[\KS h^\pm h^\mp]_D\Kpm,~(h=\pi,K)$~\cite{LHCB-PAPER-2012-027} and the ADS-like analysis of the decay mode $B^\pm\to[\Kpm\pimp\pipm\pimp]_D\Kpm$~\cite{LHCb-PAPER-2012-055}, where $[X]_D$ indicates a given final state $X$ produced by the decay of the $D$ meson. These measurements have recently been combined to yield the result $\gamma = (72.0^{+14.7}_{-15.6})^\circ$~\cite{LHCb-PAPER-2013-020}, which is in agreement with the results obtained by the \babar and \belle collaborations of $\gamma=(69^{+17}_{-16})^\circ$~\cite{Lees:2013zd} and $\gamma=(68^{+15}_{-14})^\circ$~\cite{Trabelsi:2013uj}, respectively. In analogy to studies in charged $B$ meson decays, sensitivity to \g can also be gained from decays of neutral $B$ mesons, as has been demonstrated in the LHCb analysis of $\Bz\to [\Kp\Km]_DK^{*0}$ decays~\cite{LHCB-PAPER-2012-042}.

The inclusion of additional $\Bpm\to D\Kpm$ modes can provide further constraints on $\gamma$. 
In this Letter, an analysis of the $D\to\KS\Kpm\pimp$ final states is performed, the first ADS-like analysis to use singly Cabibbo-suppressed (SCS) decays, as proposed in~\cite{Grossman:2002aq}. The two decays, $\Bpm\to[\KS\Kpm\pimp]_Dh^\pm$ and $\Bpm\to[\KS\Kmp\pipm]_Dh^\pm$, are distinguished by the charge of the $\Kpm$ from the decay of the $D$ meson relative to the charge of the $B$ meson, so that the former is labelled ``same sign'' (SS) and the latter ``opposite sign'' (OS). 

In order to interpret \CP-violating effects using these three-body decays it is necessary to account for the variation of the $D$ decay strong phase over its Dalitz plot due to the presence of resonances between the particles in the final state. Instead of employing an amplitude model to describe this phase variation, direct measurements of the phase made by the CLEO collaboration are used, which are averaged over large regions of the Dalitz plot~\cite{CLEOKsKPi}. The same CLEO study indicates that this averaging can be employed without a large loss of sensitivity. The use of the CLEO results avoids the need to introduce a systematic uncertainty resulting from an amplitude model description.

The analysis uses the full 2011 and 2012 LHCb $pp$ collision data sets, corresponding to integrated luminosities of 1 and 2\invfb and centre-of-mass energies of $\sqrt{s}=7\tev$ and $8\tev$, respectively. The results are measurements of \CP-violating observables that can be interpreted in terms of \g and other hadronic parameters of the $\Bpm$ meson decay.

\section{Formalism}
\label{sec:formalism}
The SS decay $\Bp\to[\KS\Kp\pim]_D\Kp$ can proceed via a \Dz or \Dzb meson, so that the decay amplitude is the sum of two amplitudes that interfere,
\begin{eqnarray}
A (m^2_{\KS K}, m^2_{\KS\pi}) = A_{\Dzb}(m^2_{\KS K}, m^2_{\KS\pi})  + r_B e^{i(\delta_B + \gamma)} A_{\Dz}(m^2_{\KS K}, m^2_{\KS\pi}),
\end{eqnarray}
where $A_{\{\Dz,\Dzb\}}(m^2_{\KS K}, m^2_{\KS\pi})$ are the $\Dz$ and $\Dzb$ decay amplitudes to a specific point in the $\KS\Kp\pim$ Dalitz plot. The amplitude ratio $r_B$ is $\frac{| A(\Bp\to\Dz\Kp) |}{| A(\Bp\to\Dzb\Kp) |} = 0.089\pm0.009$~\cite{LHCb-PAPER-2013-020} and $\delta_B$ is the strong phase difference between the $\Bp\to\Dz\Kp$ and $\Bp\to\Dzb\Kp$ decays. The calculation of the decay rate in a region of the Dalitz plot requires the evaluation of the integral of the interference term between the two $D$ decay amplitudes over that region. Measurements have been made by the CLEO collaboration~\cite{CLEOKsKPi}, where quantum-correlated $D$ decays are used to determine the integral of the interference term directly in the form of a ``coherence factor'', $\kappa_{\KS K\pi}$, and an average strong phase difference, $\delta_{\KS K\pi}$, as first proposed in Ref.~\cite{Atwood:2003mj}. The coherence factor can take a value between 0 and 1 and is defined through the expression
\begin{equation}
\kappa_{\KS\PK\pi}  e^{-i\delta_{\KS K\pi}} \equiv \frac{\int A^*_{\KS K^-\pi^+} (m^2_{\KS K},m^2_{K\pi}) A_{\KS K^+\pi^-} (m^2_{\KS K},m^2_{K\pi}) dm^2_{\KS K}dm^2_{K\pi}}{A^{\textrm{int.}}_{\KS K^-\pi^+}A^{\textrm{int.}}_{\KS K^+\pi^-}}, \label{eq:defCohFac}
\end{equation}
where $A^{\textrm{int.}}_{\KS\Kpm\pimp} = \int | A_{\KS\Kpm\pimp} (m^2_{\KS K},m^2_{K\pi}) |^2 dm^2_{\KS K}dm^2_{K\pi} $. This avoids the significant modelling uncertainty incurred by the determination of the strong phase difference between the \Dz and \Dzb amplitudes at each point in the Dalitz region from an amplitude model. 
The decay rates, $\Gamma$, to the four final states can therefore be expressed as
\begin{alignat}{5}
& \Gamma_{\textrm{SS, }DK}^\pm &&= z[ \quad && \hspace{0.1cm}1 &&+ \hspace{0.25cm}r_{B}^2r_D^2 &&+ 2 r_{B}r_D\kappa_{\KS\PK\Ppi}\cos(\delta_{B} \pm \gamma - \delta_{\KS K\pi}) \quad ]\nonumber\\
& \Gamma_{\textrm{OS, }DK}^\pm &&= z[ && \hspace{0.cm} r_{B}^2 &&+ \hspace{0.5cm}r_D^2 &&+ 2 r_{B}r_D\kappa_{\KS\PK\Ppi}\cos(\delta_{B} \pm \gamma + \delta_{\KS K\pi})\quad]
\label{eqn:defineYieldsGammetal}
\end{alignat}
where $r_D$ is the amplitude ratio for $\Dz\to\KS\Kp\pim$ with respect to $\Dz\to\KS\Km\pip$ and $z$ is the normalisation factor. Analogous equations can be written for the $\Bpm\to D\pipm$ system, with $r_{B}^{\pi}$ and $\delta_{B}^{\pi}$ replacing $r_B$ and $\delta_B$, respectively. Less interference is expected in the $\Bpm\to D\pipm$ system where the value of $r_{B}^{\pi}$ is much lower, approximately $0.015$~\cite{LHCb-PAPER-2013-020}. These expressions receive small corrections from mixing in the charm system which, though accounted for in Sect.~\ref{sec:Interpretation}, are not explicitly written here. At the current level of precision these corrections have a negligible effect on the final results.

Observables constructed using the decay rates in Eq.~(\ref{eqn:defineYieldsGammetal}) have a sensitivity to $\gamma$ that depends upon the value of the coherence factor, with a higher coherence corresponding to greater sensitivity. The CLEO collaboration measured the coherence factor and average strong phase difference in two regions of the Dalitz plot: firstly across the whole Dalitz plot, and secondly within a region $\pm 100\mevcc$ around the $K^*(892)^\pm$ resonance, which decays to $\KS\pipm$, where, though the sample size is diminished, the coherence is higher. The measured values are $\kappa_{\KS K\pi} = 0.73\pm0.08$ and $\delta_{\KS K\pi} = 8.3\pm15.2^\circ$ for the whole Dalitz plot, and $\kappa_{\KS K\pi} = 1.00\pm0.16$ and $\delta_{\KS K\pi} = 26.5\pm15.8^\circ$ in the restricted region. The branching fraction ratio of $\Dz\to\KS\Kp\pim$ to $\Dz\to\KS\Km\pip$ decays was found to be $0.592\pm0.044$ in the whole Dalitz plot and $0.356\pm0.034$ in the restricted region~\cite{CLEOKsKPi}.

Eight yields are measured in this analysis, from which seven observables are constructed. The charge asymmetry is measured in each of the four decay modes, defined as $\mathcal{A}_{\textrm{SS, } DK} \equiv \frac{N^{DK^-}_{\textrm{SS}}-N^{DK^+}_{\textrm{SS}}}{N^{DK^-}_{\textrm{SS}}+N^{DK^+}_{\textrm{SS}}}$ for the $\Bpm\to[\KS\Kpm\pimp]_D\Kpm$ mode and analogously for the other modes. The ratios of $\Bpm\to D\Kpm$ and $\Bpm\to D\pipm$ yields are determined for the SS and OS decays, $\mathcal{R}_{DK/D\pi\textrm{, SS}}$ and $\mathcal{R}_{DK/D\pi\textrm{, OS}}$ respectively, and the ratio of SS to OS $\Bpm\to D\pipm$ yields, $\mathcal{R}_{\textrm{SS/OS}}$, is measured. The measurements are performed both for the whole $D$ Dalitz plot and in the restricted region around the $K^*(892)^\pm$ resonance.

\section{The LHCb detector and data set}
\label{sec:Detector}
The \lhcb detector~\cite{Alves:2008zz} is a single-arm forward
spectrometer covering the \mbox{pseudorapidity} range $2<\eta <5$,
designed for the study of particles containing \bquark or \cquark
quarks. The detector includes a high-precision tracking system
consisting of a silicon-strip vertex detector surrounding the $pp$
interaction region, a large-area silicon-strip detector located
upstream of a dipole magnet with a bending power of about
$4{\rm\,Tm}$, and three stations of silicon-strip detectors and straw
drift tubes placed downstream.
The combined tracking system provides a momentum measurement with
relative uncertainty that varies from 0.4\,\% at 5\gevc to 0.6\,\% at 100\gevc,
and impact parameter (IP) resolution of 20\mum for
tracks with large transverse momentum. Different types of charged hadrons are distinguished by particle identification (PID) information
from two ring-imaging Cherenkov (RICH) detectors~\cite{LHCb-DP-2012-003}. Photon, electron and
hadron candidates are identified by a calorimeter system consisting of
scintillating-pad and preshower detectors, an electromagnetic
calorimeter and a hadronic calorimeter. Muons are identified by a
system composed of alternating layers of iron and multiwire
proportional chambers.

The trigger consists of a hardware stage, based on information from the calorimeter and muon
systems, followed by a software stage, which applies a full event
reconstruction.
The software trigger searches for a track with large \pt and large IP with respect to any $pp$ interaction point, also called a primary vertex (PV). This track is then required to be part of a two-, three- or four-track secondary vertex with a high \pt sum, significantly displaced from any PV. A multivariate algorithm~\cite{BBDT} is used for the identification of secondary vertices consistent with the decay of a \bquark hadron. 

Samples of around two million $\Bpm\to[\KS\Kmp\pipm]_D\pipm$ and two million $\Bpm\to[\KS\Kmp\pipm]_D\Kpm$ decays are simulated to be used in the analysis, along with similarly-sized samples of $\Bpm\to[\KS\pip\pim]_D\pipm$, $\Bpm\to[\KS\Kp\Km]_D\pipm$  and $\Bpm\to[\Kpm\pimp\pip\pim]_D\pipm$ decays that are used to study potential backgrounds. In the simulation, $pp$ collisions are generated using
\pythia~\cite{Sjostrand:2006za,*Sjostrand:2007gs} 
 with a specific \lhcb
configuration~\cite{LHCb-PROC-2010-056}.  Decays of hadronic particles
are described by \evtgen~\cite{Lange:2001uf}, in which final state
radiation is generated using \photos~\cite{Golonka:2005pn}. The
interaction of the generated particles with the detector and its
response are implemented using the \geant
toolkit~\cite{Allison:2006ve, *Agostinelli:2002hh} as described in
Ref.~\cite{LHCb-PROC-2011-006}.

\section{Candidate selection}
\label{sec:selection}
Candidate $B\to[\KS\Kpm\pimp]_DK$ and $B\to[\KS\Kpm\pimp]_D\pi$ decays are reconstructed in events selected by the trigger and then the candidate momenta are refit, constraining the masses of the neutral $D$ and \KS mesons to their known values~\cite{PDG2012} and the \Bpm meson to originate from the primary vertex~\cite{Hulsbergen:2005pu}. Candidates where the \KS decay is reconstructed using ``long'' pion tracks, which leave hits in the VELO and downstream tracking stations, are analysed separately from those reconstructed using ``downstream'' pion tracks, which only leave hits in tracking stations beyond the VELO. The signal candidates in the former category are reconstructed with a better invariant mass resolution.

The reconstructed masses of the $D$ and \KS mesons are required to be within 25\mevcc and 15\mevcc, respectively, of their known values. Candidate $\Bpm\to D\Kpm$ decays are separated from $\Bpm\to D\pipm$ decays by using PID information from the RICH detectors.  A boosted decision tree (BDT)~\cite{Breiman,AdaBoost} that has been developed for the analysis of the  topologically similar decay mode $\Bpm\to[\KS h^+h^-]_Dh'^\pm$ is applied to the reconstructed candidates. 
The BDT was trained using simulated signal decays, generated uniformly over the \Dz Dalitz plot, and background candidates taken from the \Bpm invariant mass region in data between 5700 and 7000\mevcc.  It exploits the displacement of tracks from the decays of long-lived particles with respect to the PV through the use of $\chi^2_{\textrm{IP}}$ variables, where $\chi^2_{\textrm{IP}}$ is defined as the difference in $\chi^2$ of a given PV fit with and without the considered track. The BDT also employs the $\Bpm$ and $D$ candidate momenta, an isolation variable sensitive to the separation of the tracks used to construct the $\Bpm$ candidate from other tracks in the event, and the $\chi^2$ per degree of freedom of the decay refit. 
In addition to the requirement placed on the BDT response variable, each composite candidate is required to have a vector displacement of production and decay vertices that aligns closely to its reconstructed momenta. The cosine of the angle between the displacement and momentum vectors is required to be less than 0.142\rad for the \KS and \Dz candidates, and less than 0.0141\rad (0.0100\rad) for long (downstream) \Bpm candidates.

Additional requirements are used to suppress backgrounds from specific processes. Contamination from $B$ decays that do not contain an intermediate $D$ meson is minimised by placing a minimum threshold of 0.2\ps on the decay time of the $D$ candidate. A potential background could arise from processes where a pion is misidentified as a kaon or vice versa. One example is the relatively abundant mode $\Bpm\to[\KS\pip\pim]_Dh^\pm$, which has a branching fraction around ten times larger than the signal. These are suppressed by placing requirements on both the $D$ daughter pion and kaon, making use of PID information. For \KS candidates formed from long tracks, the flight distance $\chi^2$ of the candidate is used to suppress background from $\Bpm\to[\Kpm\pimp\pip\pim]_D h^\pm$ decays. Where multiple candidates are found belonging to the same event, the candidate with the lowest value of the refit $\chi^2$ per degree of freedom is retained and any others are discarded, leading to a reduction in the sample size of approximately 0.3\,\%. 

\begin{figure}[p!]
\centering
SS candidates\\
\includegraphics[width=.45\textwidth]{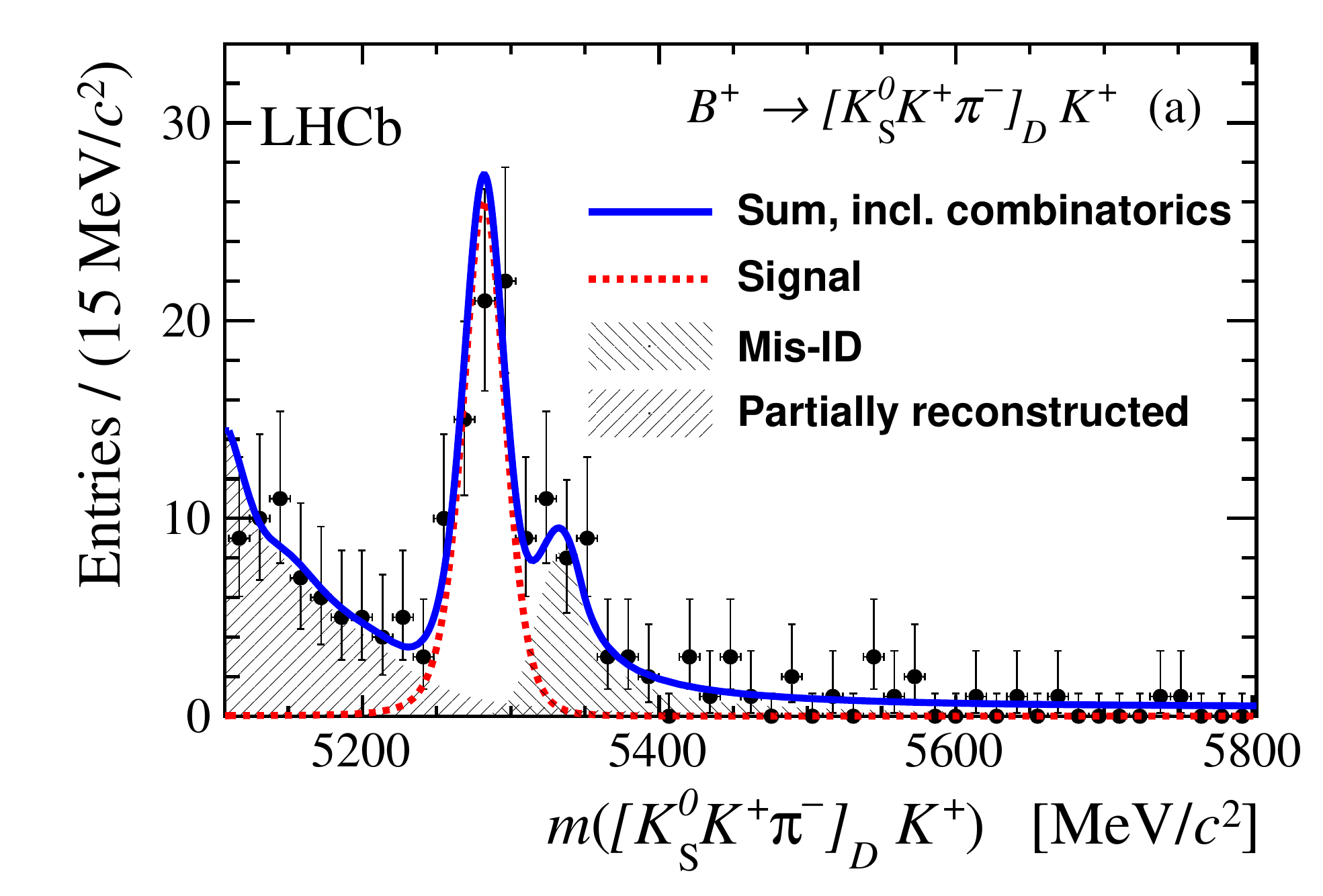}
\includegraphics[width=.45\textwidth]{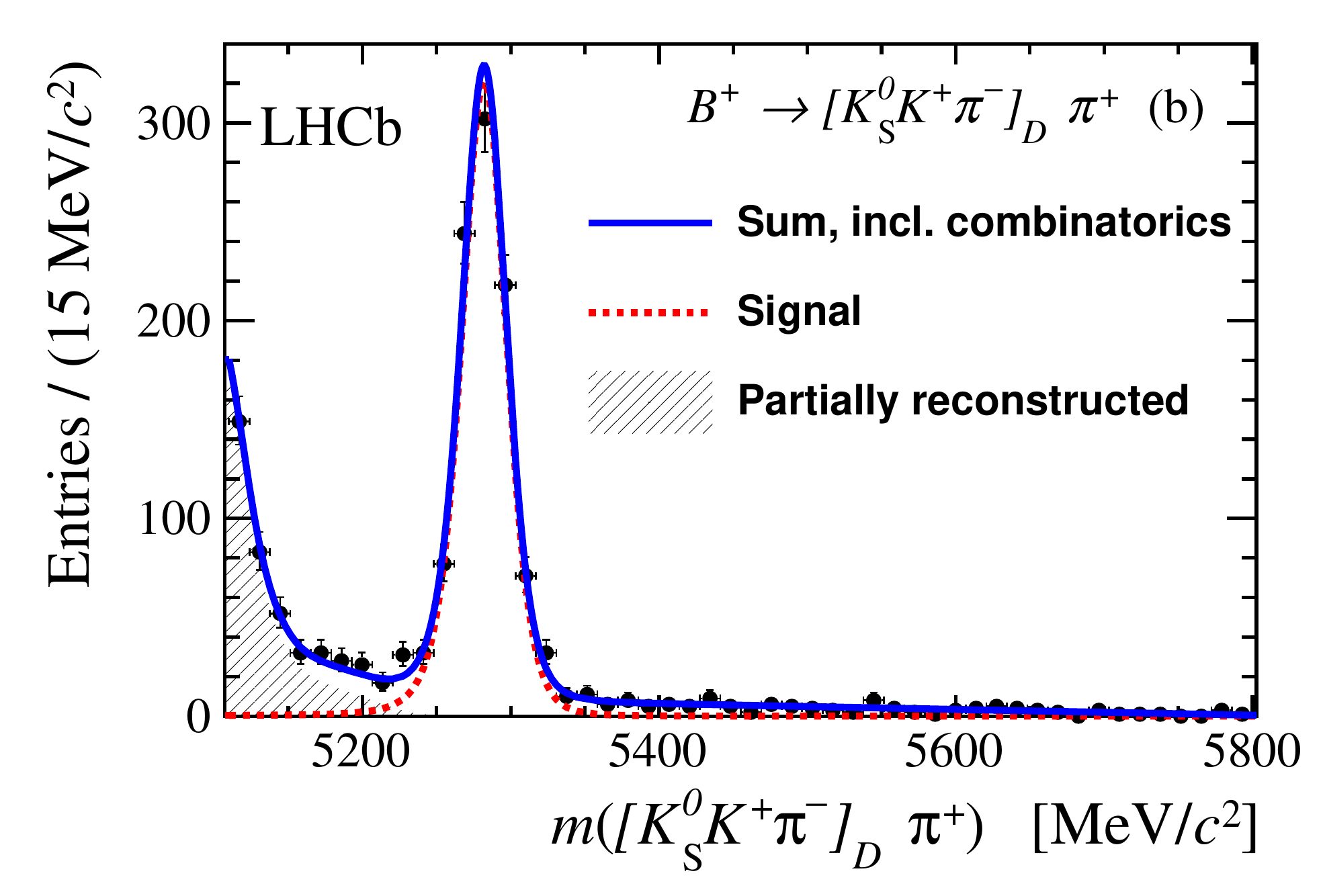}\\
\includegraphics[width=.45\textwidth]{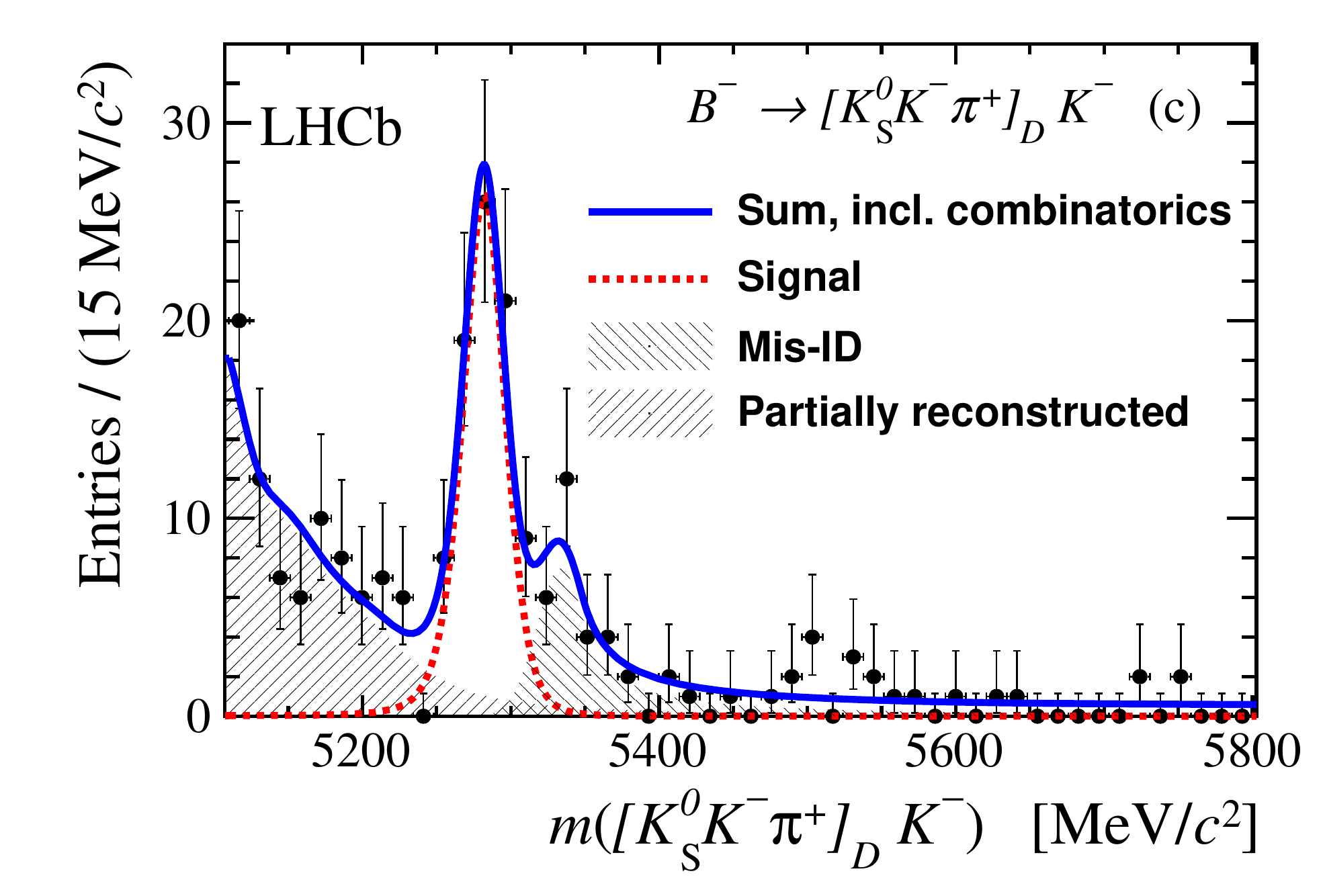}
\includegraphics[width=.45\textwidth]{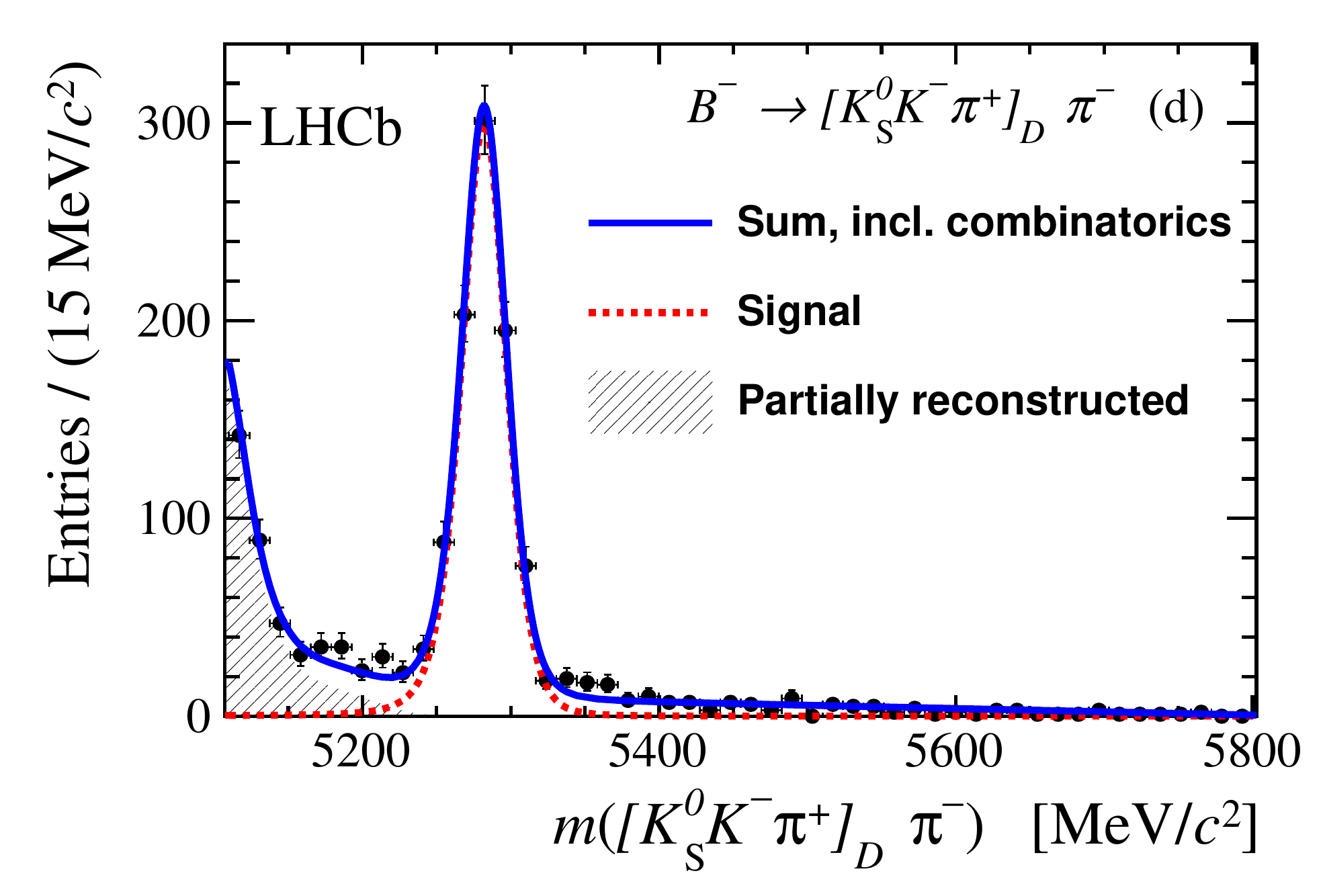}\\
OS candidates\\
\includegraphics[width=.45\textwidth]{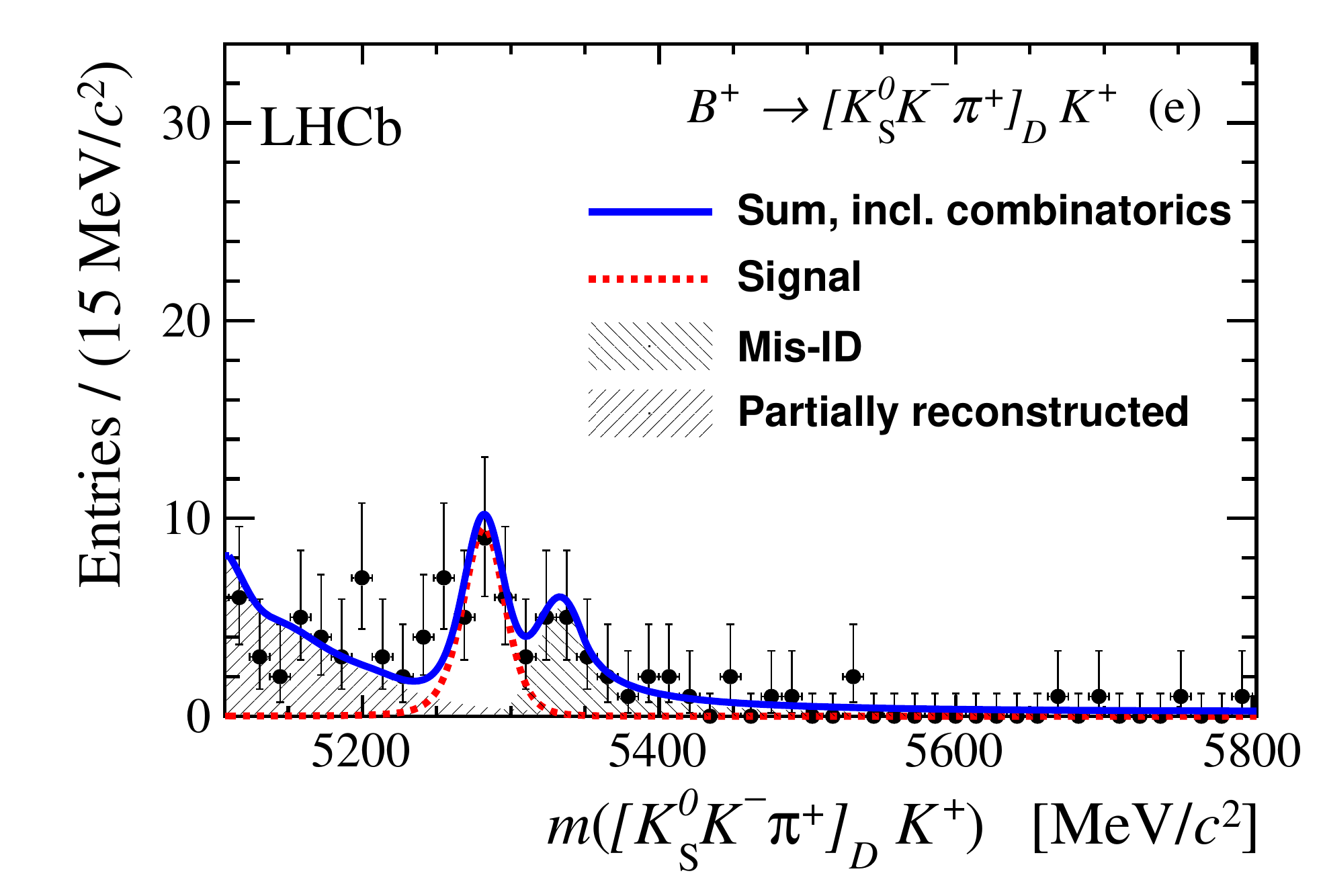}
\includegraphics[width=.45\textwidth]{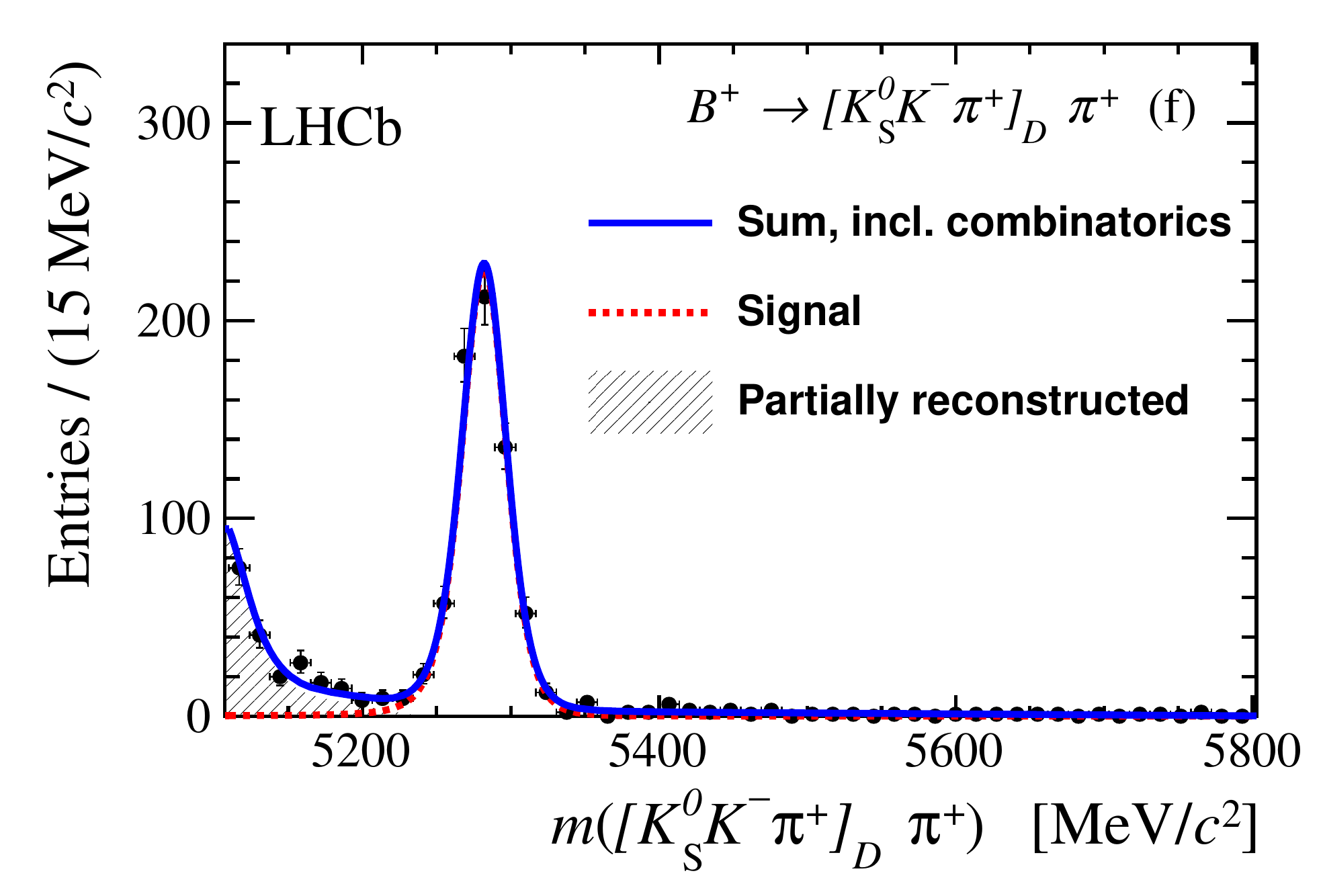}\\
\includegraphics[width=.45\textwidth]{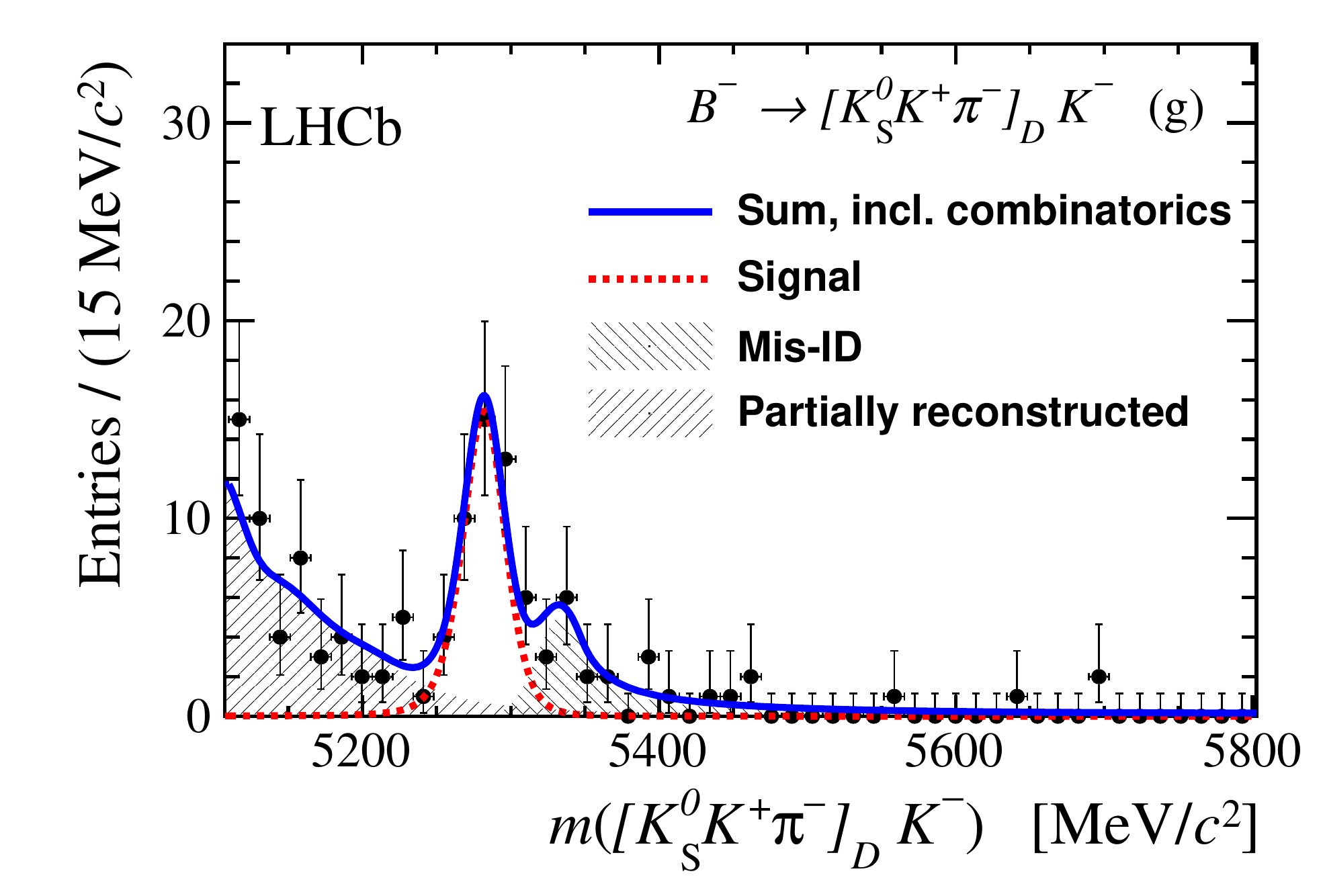}
\includegraphics[width=.45\textwidth]{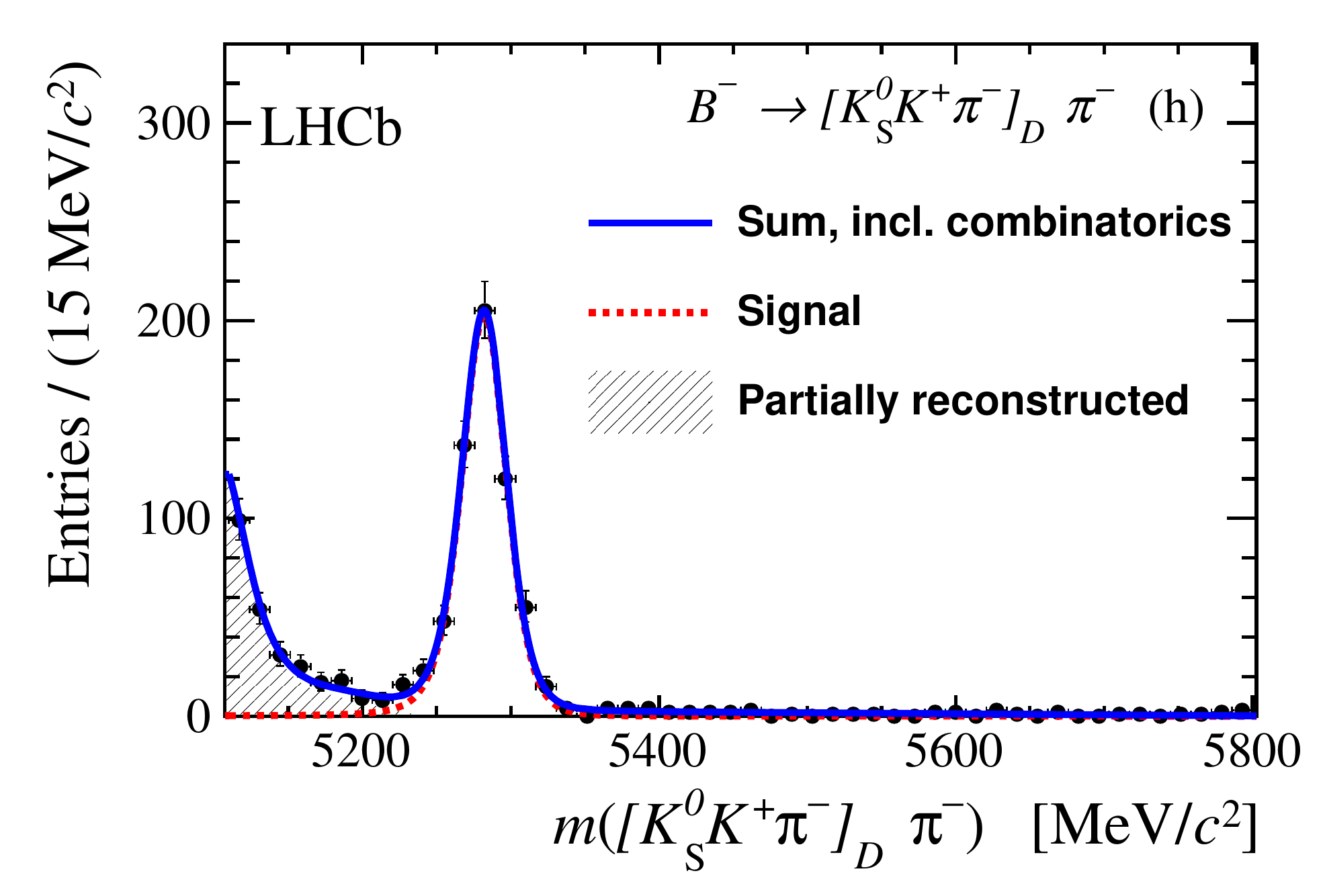}\\
\caption{Distributions of \Bpm invariant mass of the SS and OS samples for the (a, c, e, g) $\Bpm\to D\Kpm$ and (b, d, f, h) $\Bpm\to D\pipm$ candidates in the full data sample. The fits are shown for (a, b, e, f) \Bp and (c, d, g, h) \Bm candidates. Fit PDFs are superimposed.
\label{im:MassFit}}
\end{figure}

\begin{figure}[htbp]
\centering
\includegraphics[width=.49\textwidth]{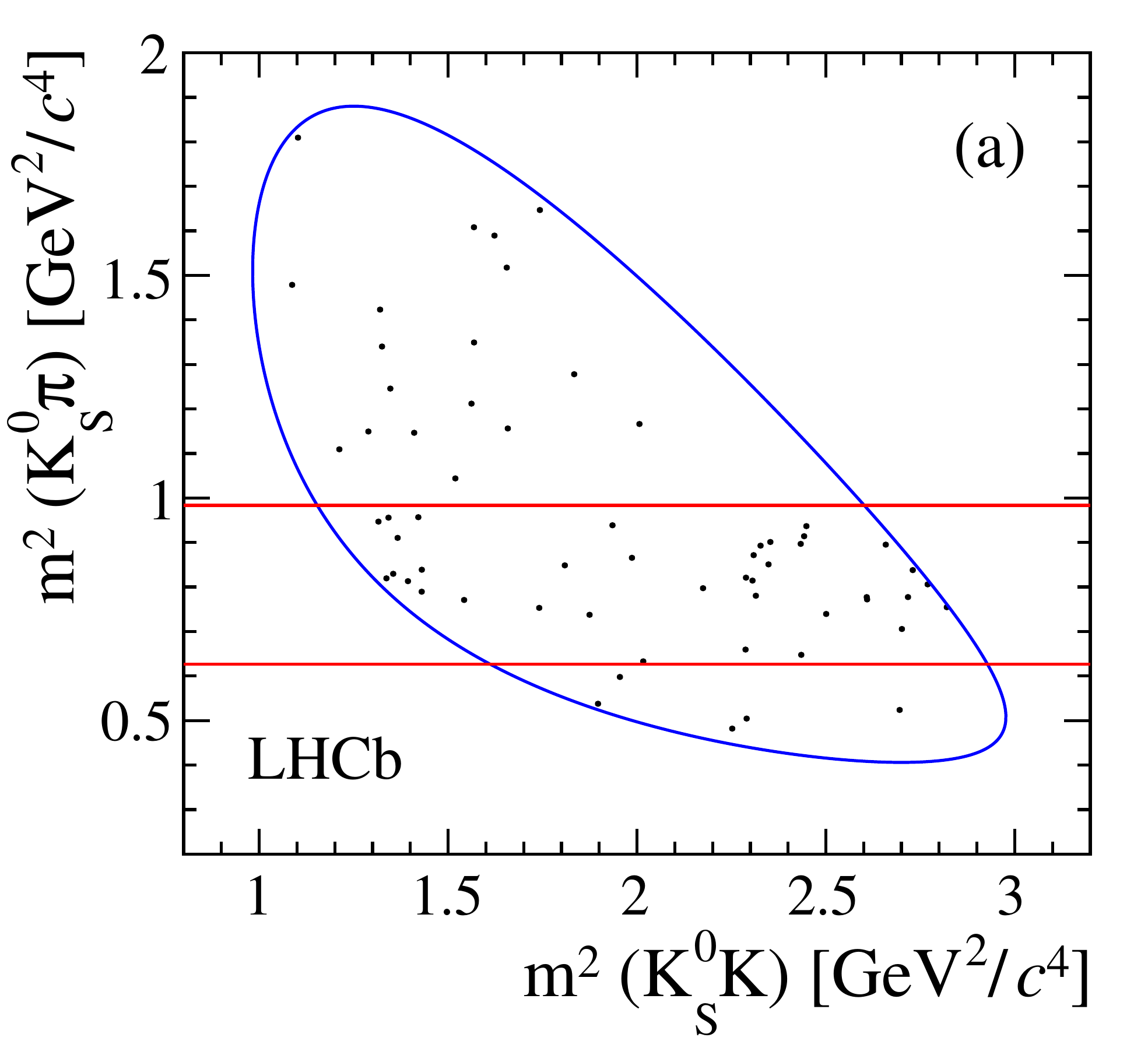}
\includegraphics[width=.49\textwidth]{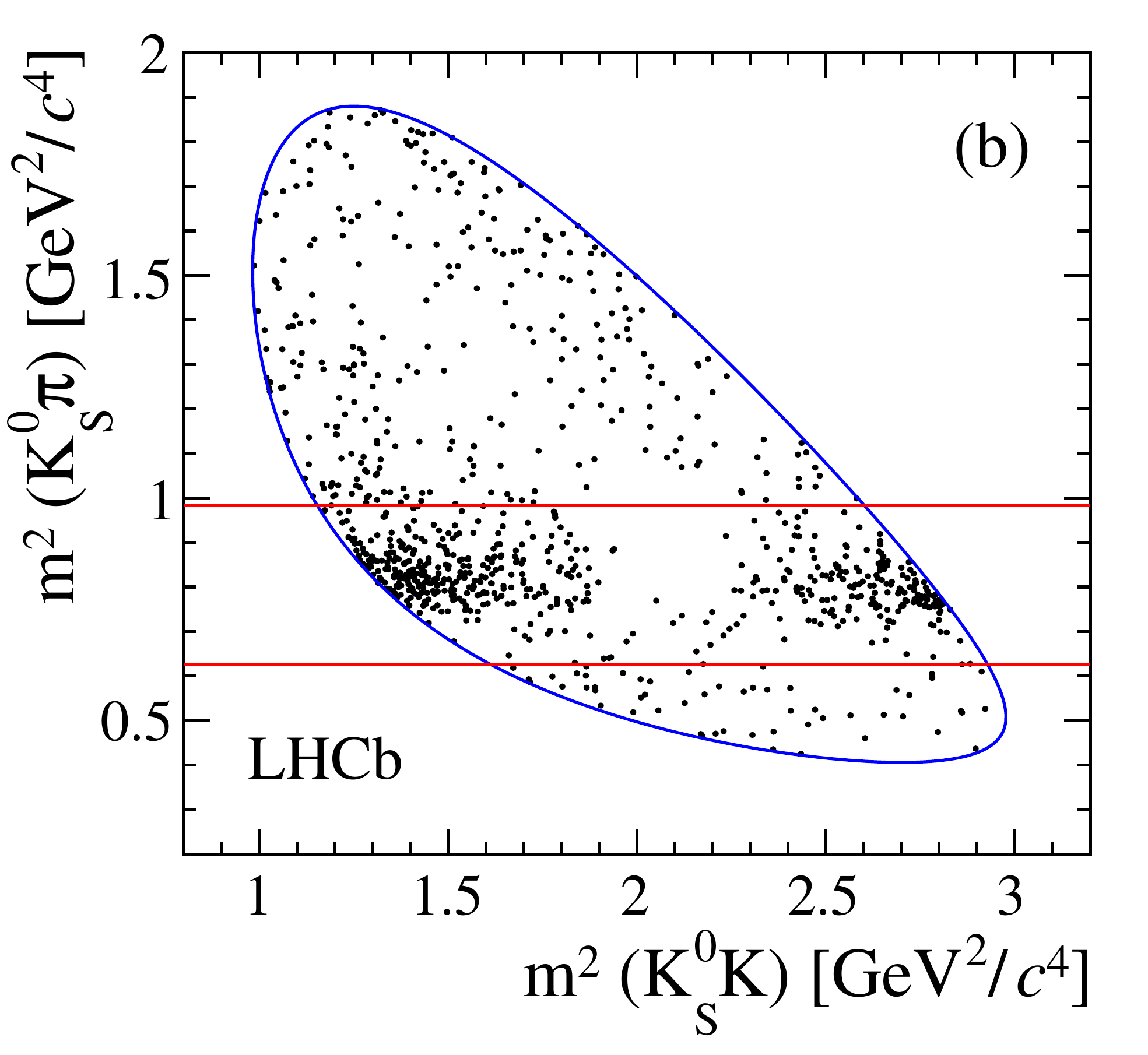}
\caption{Dalitz plot distribution of candidates selected in (a) the $\Bpm\to[\KS K\pi]_D\Kpm$ and (b) the $\Bpm\to[\KS K\pi]_D\pipm$ decay modes, where the data in the SS and OS modes, and the two \KS categories, are combined. Candidates included are required to have a refitted $\Bpm$ mass in a nominal signal window between 5247\mevcc and 5317\mevcc. The kinematic boundary is added in blue, and the restricted region around the $K^*(892)^\pm$ resonance is indicated by horizontal red lines. \label{im:DataDalitz} }
\end{figure}

The $\Bpm$ invariant mass spectra are shown in Fig.~\ref{im:MassFit} for candidates selected in the whole $D$ Dalitz plot, overlaid with a parametric fit described in Sect.~\ref{sec:invMassFit}. The $D$ Dalitz plots are shown in Fig.~\ref{im:DataDalitz} for the $\Bpm\to D\Kpm$ and $\Bpm\to D\pipm$ candidates that fall within a nominal $\Bpm$ signal region in $\Bpm$ invariant mass (5247--5317$\mevcc$). The dominant $K^*(892)^\pm$ resonance is clearly visible within a horizontal band, and the window around this resonance used in the analysis is indicated.

\section{Invariant mass fit}
\label{sec:invMassFit}
In order to determine the signal yields in each decay mode, simultaneous fits are performed to the \Bpm invariant mass spectra in the range 5110\mevcc to 5800\mevcc in the different modes, both for candidates in the whole $D$ Dalitz plot, and for only those inside the restricted region around the $K^*(892)^\pm$ resonance. The data samples are split according to the year in which the data were taken, the decay mode, the \KS type and the charge of the $B$ candidate. The fit is parameterised in terms of the observables described in Sect.~\ref{sec:formalism}, rather than varying each signal yield in each category independently. 

The probability density function (PDF) used to model the signal component is a modified Gaussian function with asymmetric tails, where the unnormalised form is given by
\begin{equation}
f(m; m_0,\alpha_L,\alpha_R,\sigma) \equiv \left\lbrace{\exp[-(m-m_0)^2/(2\sigma^2 + \alpha_L(m-m_0)^2)] \textrm{ for } m<m_0, \atop \exp[-(m-m_0)^2/(2\sigma^2 + \alpha_R(m-m_0)^2)] \textrm{ for } m>m_0,}\right.
\end{equation}
where $m$ is the reconstructed mass, $m_0$ is the mean $B$ mass and $\sigma$ determines the width of the function. The $\alpha_{L,R}$ parameters govern the shape of the tail. The mean $B$ mass is shared among all categories but is allowed to differ according to the year in which the data were collected. The $\alpha_L$ parameters are fixed to the values determined in the earlier analysis of $\Bpm\to[\KS\pip\pim]_D h^\pm$~\cite{LHCB-PAPER-2012-027}. The $\alpha_R$ parameters are common to the $\Bpm\to D\pipm$ and $\Bpm\to D\Kpm$, SS and OS categories, and are allowed to vary in the fit. Only the width parameters $\sigma(\Bpm \to D\Kpm)$ are allowed to vary in the fit. The ratios $\sigma(\Bpm\to D\pipm)/\sigma(\Bpm\to D\Kpm)$ are fixed according to studies of the similar mode $\Bpm\to[\KS\pip\pim]_Dh^\pm$. The fitted values for $\sigma(\Bpm\to D\Kpm)$ vary by less than 10\% around 14\mevcc. 
The total yield of $\Bpm\to D\pipm$ decays is allowed to vary between the different \KS type and year categories. The yields in the various $D$ decay modes and different charges, and all the $\Bpm\to D\Kpm$ yields, are determined using the observables described in Sect.~\ref{sec:formalism}, rather than being fitted directly.

In addition to the signal PDF, two background PDFs are required. The first background PDF models candidates formed from random combinations of tracks and is represented by a linear function.  
In the fit within the restricted Dalitz region, where the sample size is significantly smaller, the slope of the linear function fitting the $\Bpm\to D\pipm$ data is fixed to the value determined in the fit to the whole Dalitz plot. The second background PDF accounts for contamination from partially reconstructed processes. Given that the contamination is dominated by those processes that involve a real $\Dz\to\KS\Kpm\pimp$ decay, the PDF is fixed to the shape determined from the more abundant mode $\Bpm\to[\Kpm\pimp]_D h^\pm$. The yields of both these background components are free to vary in each data category.

A further significant background is present in the $\Bpm\to D\Kpm$ samples due to $\pi\to K$ misidentification of the much more abundant $\Bpm\to D\pipm$ mode. This background is modelled in the $\Bpm\to D\Kpm$ spectrum using a Crystal Ball function~\cite{Skwarnicki:1986xj}, where the parameters of the function are common to all data categories in the fit and are allowed to vary. The yield of the background in the $\Bpm\to D\Kpm$ samples is fixed with respect to the fitted $\Bpm\to D\pipm$ signal yield using knowledge of the RICH particle identification efficiencies that is obtained from data using samples of $D^{*\pm}\to [K\pi]_D\pipm$ decays. The efficiency for kaons to be selected is found to be around 84\,\% and the misidentification rate for pions is around 4\,\%.

Production and detection asymmetries are accounted for, following the same procedure as in Refs.~\cite{LHCb-PAPER-2012-001,LHCb-PAPER-2012-055}. Values for the \Bpm production and $K$ detection asymmetries are assigned such that the combination of production and detection asymmetries corresponds to the raw asymmetry observed in $\Bpm\to J/\psi \Kpm$ decays~\cite{LHCb-PAPER-2011-024}. The detection asymmetry assigned is $-0.5\pm0.7\,\%$ for each unit of strangeness in the final state to account for the differing interactions of \Kp and \Km mesons with the detector material. An analogous asymmetry is present for pions, though it is expected to be much smaller, and the detection asymmetry assigned is $0.0\pm0.7\,\%$. Any potential asymmetry arising from a difference between the responses of the left and right sides of the detector is minimised by combining approximately equal data sets taken with opposite magnet polarity.

\begin{figure}[htbp]
\centering
\includegraphics[width=.6\textwidth]{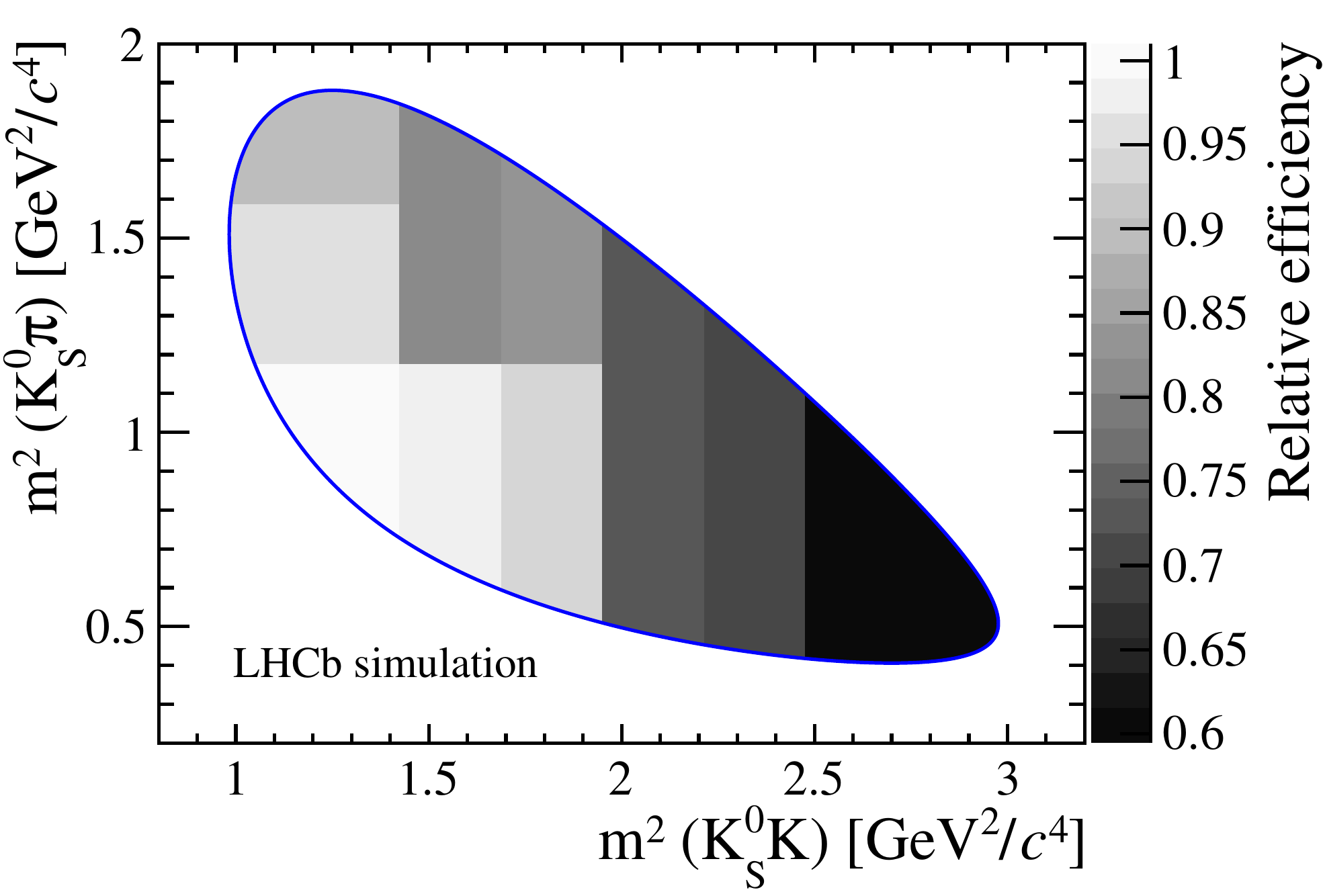}
\caption{Dalitz acceptance determined using simulated events and normalised relative to the maximum efficiency. \label{im:DalitzAcceptance}} 
\end{figure}

A further correction is included to account for non-uniformities in the acceptance over the Dalitz plot. 
This efficiency correction primarily affects the $\mathcal{R}_{\textrm{SS/OS}}$ observable, given the difference in the Dalitz distributions for the two $D$ meson decay modes. A correction factor, $\zeta$, is found by combining the LHCb acceptance, extracted from the simulated signal sample, and amplitude models, $A_{\textrm{SS, OS}}(m^2_{\KS K},m^2_{\KS\pi})$, for the Dalitz distributions of the SS or OS decays,
\begin{equation}
\zeta \equiv \frac{ \int_{\mathcal{D}} \textrm{d}m^2_{\KS K} \textrm{d}m^2_{\KS\pi} [\epsilon(m^2_{\KS K},m^2_{\KS\pi}) \times | A_{\textrm{OS}}(m^2_{\KS K},m^2_{\KS\pi})|^2]   }{ \int_{\mathcal{D}} \textrm{d}m^2_{\KS K} \textrm{d}m^2_{\KS\pi} [\epsilon(m^2_{\KS K},m^2_{\KS\pi}) \times | A_{\textrm{SS}}(m^2_{\KS K},m^2_{\KS\pi})|^2]    },
\label{eqn:EffCorrInteg}
\end{equation}
where $\epsilon(m^2_{\KS K},m^2_{\KS\pi})$ is the efficiency at a point in the Dalitz plot. The typical deviation of $\zeta$ from unity is found to be around 5\,\%. The acceptance is illustrated in Fig.~\ref{im:DalitzAcceptance}, where bins of variable size are used to ensure that statistical fluctuations due to the finite size of the simulated sample are negligible. The Dalitz distributions are determined using the fact that little interference is expected in $\Bpm\to D\pipm$ decays and, therefore, the flavour of the $D$ meson is effectively tagged by the charge of the pion. In this case, the Dalitz distributions are given by considering the relevant \Dz decay ($\Dz\to\KS\Km\pip$ for SS and $\Dz\to\KS\Kp\pim$ for OS). These \Dz decay Dalitz distributions are known and amplitude models from CLEO are available~\cite{CLEOKsKPi} from which the Dalitz distributions can be extracted. 

Due to the restricted sample size under study, small biases exist in the determination of the observables. The biases are determined by generating and fitting a large number of simulated samples using input values obtained from the fit to data, and are typically found to be around 2\,\%. The fit results are corrected accordingly.

The fit projections, with long and downstream \KS-type categories merged and 2011 and 2012 data combined, are given for the fit to the whole Dalitz plot in Fig.~\ref{im:MassFit}. The signal purity in a nominal mass range from $5247\mevcc$ to 5317\mevcc is around 85\,\% for the $\Bpm\to D\Kpm$ samples and 96\,\% for the $\Bpm\to D\pipm$ samples.  
The signal yields derived from the fits to both the whole and restricted region of the Dalitz plot are given in Table~\ref{tab:SignalYields}. The fitted values of the observables are given in Table~\ref{tab:AnalysisResults}, including their systematic uncertainties as discussed in Sect.~\ref{sec:Systematics}. The only significant difference between the observables fitted in the two regions is for the value of $\mathcal{R}_{\textrm{SS/OS}}$. This ratio is expected to differ significantly, given that the fraction of $\Dz\to\KS\Km\pip$ decays that are expected to lie inside the restricted portion of the Dalitz plot is around 75\,\%, whereas for $\Dz\to\KS\Kp\pim$ the fraction is around 44\,\%~\cite{CLEOKsKPi}. This accounts for the higher value of $\mathcal{R}_{\textrm{SS/OS}}$ in the restricted region. The ratios between the $\Bpm\to D\Kpm$ and $\Bpm\to D\pipm$ yields are consistent with that measured in the LHCb analysis of $\Bpm\to[K\pi]_Dh^\pm$, $0.0774 \pm 0.0012 \pm 0.0018$~\cite{LHCb-PAPER-2012-001}. The \CP asymmetries are consistent with zero in the $\Bpm\to D\pipm$ system, where the effect of interference is expected to be small.  
The asymmetries in the $\Bpm\to D\Kpm$ system, $\mathcal{A}_{\textrm{SS, }DK}$ and $\mathcal{A}_{\textrm{OS, }DK}$, which have the highest sensitivity to $\gamma$ are all compatible with zero at the $2\sigma$ level. 
The correlations between $\mathcal{R}_{\textrm{SS/OS}}$ ratio and the ratios $\mathcal{R}_{DK/D\pi\textrm{, SS}}$ and $\mathcal{R}_{DK/D\pi\textrm{, OS}}$ are $-16\,\%$ ($-13\,\%$) and $+16\,\%$ ($+16\,\%$), respectively, for the fit to the whole Dalitz plot ($K^*(892)^\pm$ region). The correlation between the $\mathcal{R}_{DK/D\pi\textrm{, SS}}$ and $\mathcal{R}_{DK/D\pi\textrm{, OS}}$ ratios is $+11\,\%$ ($+15\,\%$). Correlations between the asymmetry observables are all less than 1\,\% and are neglected.

\begin{table}[t!]
\centering
\caption{Signal yields and their statistical uncertainties derived from the fit to the whole Dalitz plot region, and in the restricted region of phase space around the $K^*(892)^\pm$ resonance. \label{tab:SignalYields}}
\begin{tabular}{l  r@{\hspace{0.1cm}$\pm$\hspace{0.1cm}}r r@{\hspace{0.1cm}$\pm$\hspace{0.1cm}}r r@{\hspace{0.1cm}$\pm$\hspace{0.1cm}}r r@{\hspace{0.1cm}$\pm$\hspace{0.1cm}}r}
\hline
& \multicolumn{4}{c}{Whole Dalitz plot} & \multicolumn{4}{c}{$K^*(892)^\pm$ region} \\
Mode  & \multicolumn{2}{c}{$D\Kpm$} &  \multicolumn{2}{c}{$D\pipm$} &  \multicolumn{2}{c}{$D\Kpm$} &  \multicolumn{2}{c}{$D\pipm$}\\
\hline
SS & 145 & 15 	& 1841 & 47	& 97 & 12	& 1365 & 38\\
OS & 71 & 10	& 1267 & 37	& 26 & 6	& 553 & 24\\
\hline
\end{tabular}
\end{table}

\begin{table}[h!]
\centering
\caption{Results for the observables measured in the whole Dalitz plot region, and in the restricted region of phase space around the $K^*(892)^\pm$ resonance. The first uncertainty is statistical and the second is systematic. The corrections for production and detection asymmetries are applied, as is the efficiency correction defined in Eq.~(\ref{eqn:EffCorrInteg}). \label{tab:AnalysisResults}}
\begin{tabular}{l r@{\hspace{0.1cm}$\pm$\hspace{0.1cm}}l@{\hspace{0.1cm}$\pm$\hspace{0.1cm}}l r@{\hspace{0.1cm}$\pm$\hspace{0.1cm}}l@{\hspace{0.1cm}$\pm$\hspace{0.1cm}}l}
\hline
Observable & \multicolumn{3}{c}{Whole Dalitz plot} & \multicolumn{3}{c}{$K^*(892)^\pm$ region} \\
\hline
$\mathcal{R}_{\textrm{SS/OS}}$ & 1.528 & 0.058 &0.025 & 2.57&0.13 &0.06\\ 
$\mathcal{R}_{DK/D\pi\textrm{, SS}}$& 0.092 & 0.009 &0.004 & 0.084& 0.011 &0.003  \\
$\mathcal{R}_{DK/D\pi\textrm{, OS}}$& 0.066 & 0.009 &0.002 & 0.056&0.013 &0.002\\
$\mathcal{A}_{\textrm{SS, }DK} $& 0.040 & 0.091 &0.018 & 0.026& 0.109 &0.029\\ 
$\mathcal{A}_{\textrm{OS, }DK} $& 0.233 & 0.129 &0.024 &0.336& 0.208 &0.026\\
$\mathcal{A}_{\textrm{SS, }D\pi}$ & $-0.025$ & 0.024 &0.010 & $-0.012$& 0.028 &0.010\\ 
$\mathcal{A}_{\textrm{OS, }D\pi}$ & $-0.052$ & 0.029 &0.017 & $-0.054$& 0.043 &0.017\\
\hline
\end{tabular}
\end{table}

\section{Systematic uncertainties}
\label{sec:Systematics}
The largest single source of systematic uncertainty is the knowledge of the efficiency correction factor that multiplies the $\mathcal{R}_{\textrm{SS/OS}}$ observable. 
This uncertainty has three sources: the uncertainties on the CLEO amplitude models, the granularity of the Dalitz divisions in which the acceptance is determined, and the limited size of the simulated sample available to determine the LHCb acceptance. Of these, it is the modelling uncertainty that is dominant.  
In addition, an uncertainty is assigned to account for the fact that interference is neglected in the computation of the efficiency correction factor, which is shared between the $D\pipm$ and $D\Kpm$ systems.

Uncertainties on the parameters that are fixed in the PDF are propagated to the observables by repeating the fit to data whilst varying each fixed parameter according to its uncertainty. 
An additional systematic uncertainty is calculated for the fit to the restricted $K^*(892)^\pm$ region, where the $D\pipm$ combinatorial background slopes are fixed to the values determined in the fit to the whole Dalitz plot.
 
Uncertainties are assigned to account for the errors on the \Bpm production asymmetry and the \Kpm and \pipm detection asymmetries. 
The effect of the detection asymmetry depends on the pion and kaon content of the final state, and the resulting systematic uncertainty is largest for the $\mathcal{A}_{\textrm{SS, }DK}$ and $\mathcal{A}_{\textrm{OS, }D\pi}$ observables.

The absolute uncertainties on the particle identification efficiencies are small, typically around 0.3\,\% for kaon efficiencies and 0.03\,\% for pion efficiencies. Of the four main sources of systematic error, these result in the smallest uncertainties on the experimental observables.

In Table~\ref{tab:systematicsSummary_WholeDalitzSpace}, the sources of systematic uncertainty are given for each observable in the fit to the whole Dalitz plot. Similarly those for the fit in the restricted region are given in Table~\ref{tab:systematicsSummary_Kstar}.

\begin{table}[ht]
\caption{Absolute values of systematic uncertainties, in units of $10^{-2}$, for the fit to the whole Dalitz plot. \label{tab:systematicsSummary_WholeDalitzSpace}}
\begin{tabular}{l l l p{2.5cm} l l}
\hline
 Observable & Eff. correction & Fit PDFs & Prod. and det. asymms. & PID  & Total\\
 \hline
 $\mathcal{R}_{\textrm{SS/OS}}$ & 2.40 & 0.50 & $-$ & 0.01 & 2.45 \\
$\mathcal{R}_{DK/D\pi\textrm{, SS}}$ & 0.01 & 0.38 & $-$ & 0.02 & 0.38 \\
$\mathcal{R}_{DK/D\pi\textrm{, OS}}$ & 0.01 & 0.19 & $-$ & 0.01 & 0.19 \\
$\mathcal{A}_{\textrm{SS, }DK}$ & 0.14 & 0.44 & 1.71 & 0.01 & 1.78 \\
$\mathcal{A}_{\textrm{OS, }DK}$ & 0.36 & 2.13 & 0.99 & 0.01 & 2.37 \\
$\mathcal{A}_{\textrm{SS, }D\pi}$ & 0.02 & 0.05 & 0.99 & $<0.01$ & 0.99 \\
$\mathcal{A}_{\textrm{OS, }D\pi}$ & 0.03 & 0.10 & 1.71 & $<0.01$ & 1.72 \\
\hline
\end{tabular}
\end{table}

\begin{table}[ht]
\caption{Absolute values of systematic uncertainties, in units of $10^{-2}$, for the fit in the restricted region. \label{tab:systematicsSummary_Kstar}}
\begin{tabular}{l l l p{2.5cm} l l}
\hline
 Observable & Eff. correction & Fit PDFs & Prod. and det. asymms. & PID & Total\\
 \hline
 $\mathcal{R}_{\textrm{SS/OS}}$ & 6.08 & 0.53 & $-$ & 0.01 & 6.10 \\
$\mathcal{R}_{DK/D\pi\textrm{, SS}}$ & 0.01 & 0.25 & $-$ & 0.02 & 0.25 \\
$\mathcal{R}_{DK/D\pi\textrm{, OS}}$ & 0.01 & 0.21 & $-$ & 0.01 & 0.21 \\
$\mathcal{A}_{\textrm{SS, }DK}$ & 0.13 & 2.27 & 1.71 & 0.01 & 2.85 \\
$\mathcal{A}_{\textrm{OS, }DK}$ & 0.04 & 2.38 & 0.99 & 0.01 & 2.57 \\
$\mathcal{A}_{\textrm{SS, }D\pi}$ & 0.04 & 0.17 & 0.99 & $<0.01$ & 1.00 \\
$\mathcal{A}_{\textrm{OS, }D\pi}$ & 0.06 & 0.09 & 1.71 & $<0.01$ & 1.72 \\
\hline
\end{tabular}
\end{table}

\section{Interpretation and conclusions}
\label{sec:Interpretation}
\begin{figure}[t!]
\centering
\includegraphics[width=.49\textwidth]{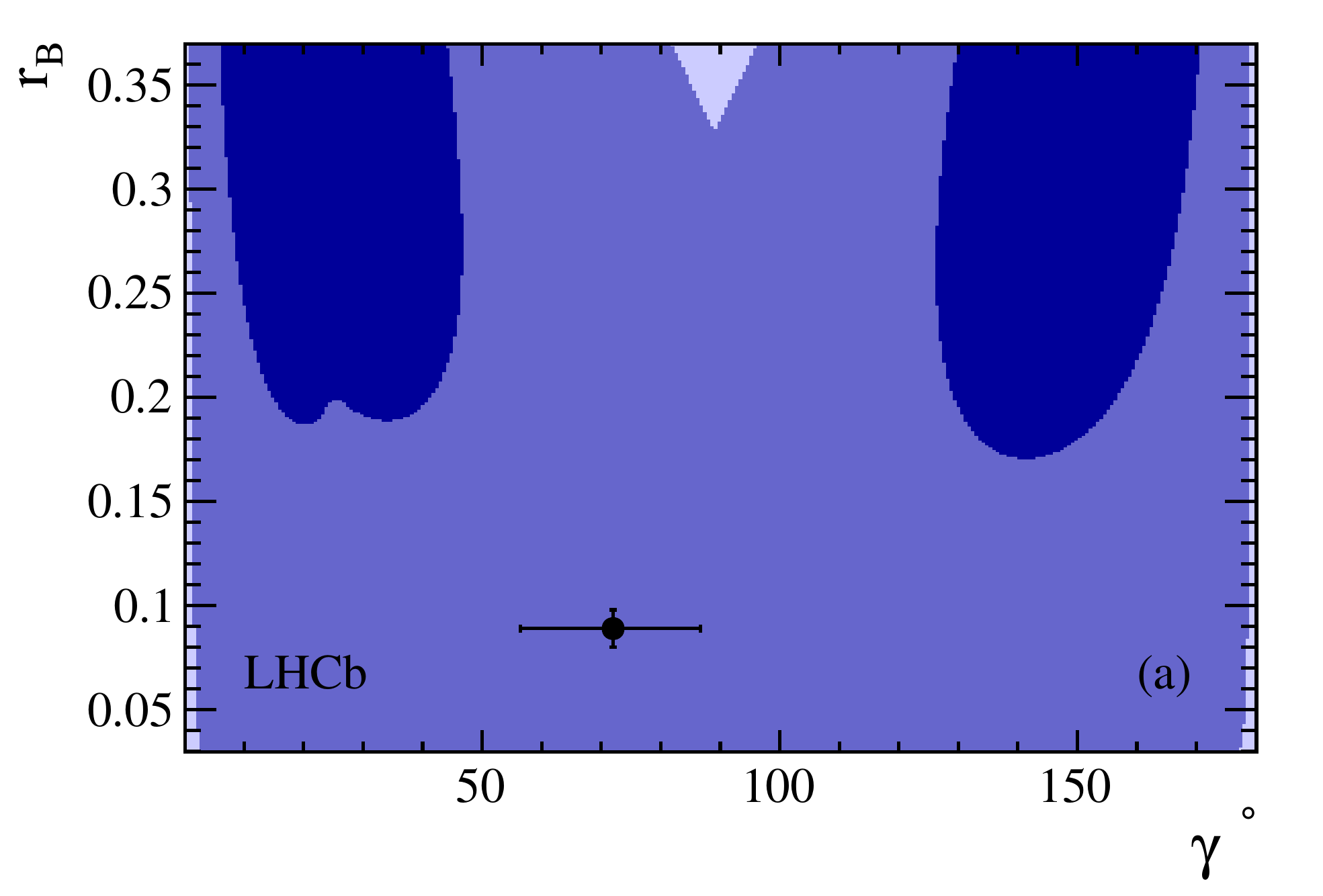}\label{im:GammaRbContours_WholeSpace}
\includegraphics[width=.49\textwidth]{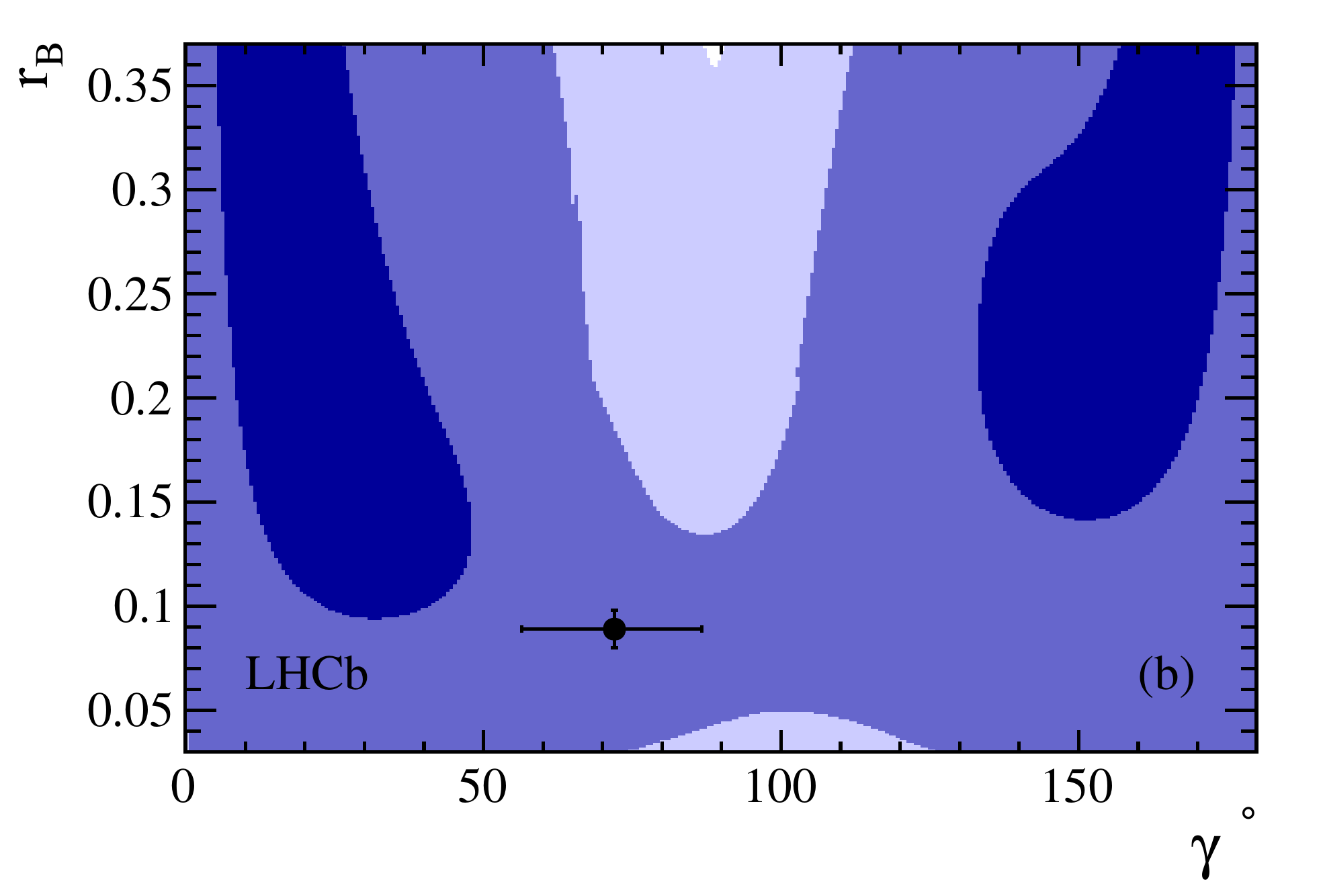}\label{im:GammaRbContours_Kstar}
\caption{Scans of the $\chi^2$ probabilities over the $\gamma-r_B$ parameter space for (a) the whole Dalitz fit and (b) the fit inside the $K^*$ region (b). The contours are the usual $n\sigma$ profile likelihood contours, where $\Delta\chi^2 = n^2$ with $n = 1 \textrm{ (dark blue), } 2 \textrm{ (medium blue), and } 3 \textrm{ (light blue)}$.  
The $2\sigma$ contour encloses almost all of the parameter space shown, so a central value of $\gamma$ and relevant bounds are not extracted. The result is seen to be compatible with the current LHCb measurement of $\gamma$, indicated by the point  at ($\gamma=72.0^\circ$ and $r_B=0.089$), at a level between 1 and $2\sigma$. \label{im:GammaRbContours} }
\end{figure}

The sensitivity of this result to the CKM angle $\gamma$ is investigated by employing a frequentist method to scan the $\gamma-r_B$ parameter space and calculate the $\chi^2$ probability at each point, given the measurements of the observables with their statistical and systematic uncertainties combined in quadrature, accounting for correlations between the statistical uncertainties. The effects of charm mixing are accounted for, but \CP violation in the decays of $D$ mesons is neglected. Regions of $1\sigma$, $2\sigma$ and $3\sigma$ compatibility with the measurements made are indicated by the dark, medium and light blue regions, respectively, in Fig.~\ref{im:GammaRbContours}. The small sample size in the current data set results in a bound on $\gamma$ that is only closed for the $1\sigma$ contour.

Although it is not possible to measure $\gamma$ directly using these results alone, it is  of interest to consider how this result relates to the previous LHCb $\gamma$ determination, obtained from other $\Bpm\to D\Kpm$ modes~\cite{LHCb-PAPER-2013-020}, since it will be included in future combinations. In order to aid this comparison, the scans of the $\gamma-r_B$ space plots are shown in Fig.~\ref{im:GammaRbContours}(a) for the measurement made using the whole $D\to\KS K\pi$ Dalitz plot and in Fig.~\ref{im:GammaRbContours}(b) for that made in the restricted region. The current LHCb average, extracted from a combination of $\Bpm\to D\Kpm$ analyses~\cite{LHCb-PAPER-2013-020}, is shown as a point with error bars at $\gamma=72.0^\circ$ and $r_B=0.089$. The LHCb average lies within the  $2\sigma$ region allowed by the measurements presented in this Letter. It is interesting to note that the bound determined in the $\gamma-r_B$ space indicates a more stringent constraint when using the restricted region, where the coherence is higher. This, and the fact that the measurements in this Letter are limited by their statistical precision, motivates the use of this region in future analyses of these decays in a larger data sample. Combination with analyses in other, more abundant channels with sensitivity to the same parameters will yield more stringent constraints upon $\gamma$.

In summary, for the first time a measurement of charge asymmetries and associated observables is presented in the decay modes $\Bpm\to[\KS\Kpm\pimp]_Dh^\pm$ and $\Bpm\to[\KS\Kmp\pipm]_Dh^\pm$, and no significant \CP violation is observed. The results of the analysis are consistent with other measurements of observables in related $\Bpm\to D\Kpm$ modes, and will be valuable in future global fits of the CKM parameter $\gamma$.

\section*{Acknowledgements}
\noindent We express our gratitude to our colleagues in the CERN
accelerator departments for the excellent performance of the LHC. We
thank the technical and administrative staff at the LHCb
institutes. We acknowledge support from CERN and from the national
agencies: CAPES, CNPq, FAPERJ and FINEP (Brazil); NSFC (China);
CNRS/IN2P3 and Region Auvergne (France); BMBF, DFG, HGF and MPG
(Germany); SFI (Ireland); INFN (Italy); FOM and NWO (The Netherlands);
SCSR (Poland); MEN/IFA (Romania); MinES, Rosatom, RFBR and NRC
``Kurchatov Institute'' (Russia); MinECo, XuntaGal and GENCAT (Spain);
SNSF and SER (Switzerland); NAS Ukraine (Ukraine); STFC (United
Kingdom); NSF (USA). We also acknowledge the support received from the
ERC under FP7. The Tier1 computing centres are supported by IN2P3
(France), KIT and BMBF (Germany), INFN (Italy), NWO and SURF (The
Netherlands), PIC (Spain), GridPP (United Kingdom).
We are indebted to the communities behind the multiple open source software packages we depend on.
We are also thankful for the computing resources and the access to software R\&D tools provided by Yandex LLC (Russia).

\addcontentsline{toc}{section}{References}
\setboolean{inbibliography}{true}
\bibliographystyle{LHCb}
\bibliography{main,LHCb-PAPER,LHCb-DP}

\end{document}